\documentclass[twocolumn,pra,eqsecnum,showpacs,amsmath
  ,superscriptaddress]{revtex4-1}
\usepackage{epsfig,graphicx}

\newcommand{\be}{\begin{equation}}
\newcommand{\e}[1]{\label{#1}\end{equation}}
\def\bea{\begin{eqnarray}}
\def\ea#1{\label{#1}\end{eqnarray}}
\def\rqn#1{(\ref{#1})}
\def\ee{\end{equation}}
\def\eea{\end{eqnarray}}
\def\bes#1{\begin{subequations}\label{#1}}
\def\ese{\end{subequations}}

\begin{document}

\title{
Multiphoton electron detachment by a superposition of static and ac fields
}
\author{A. G. Kofman}
 \affiliation{Advance Science Institute, RIKEN, Wako-shi, Saitama 351-0198, Japan}
 \affiliation{Physics Department, The University of Michigan, Ann Arbor, Michigan 48109-1040, USA}
%
\author{G. P. Berman}
\affiliation{Theoretical Division, Los Alamos National
Laboratory, Los Alamos, New Mexico 87545}
\date{\today}

\begin{abstract}
A theory of electron detachment from atoms or negative ions by a superposition
of a static and a laser (or, more generally, ac) fields, parallel to each other, is developed in the case when the photon energy and the field amplitude are much less than the detachment energy and intra-ion field, respectively.
 Simple analytical results together with their validity conditions
are obtained.
Several qualitatively different regimes of detachment have been identified.
Applications of the present theory to electron emission from metal
and semiconductor surfaces and from photosynthetic bio-complexes are discussed.
\end{abstract}

\maketitle

\section{Introduction}
\label{I}

The photoeffect by a strong nonresonant field in the presence of a
static electric field is a problem of a significant scientific interest.
 This problem can arise in different situations.
 A static field can accelerate ionization of atoms and electron detachment from negative ions in the presence of a nonresonant laser field \cite{aru70}.
 It is also of interest to consider effects of an ac field on static-field induced electron emission from metal and semiconductor surfaces and from photosynthetic bio-complexes.
 Such  experiments can be done, e.g., by a modification of the techniques of scanning tunneling microscopy \cite{sca93,Lukins1,Lukins2,Lukins3}.

Such phenomena are not completely understood yet, in spite of the fact they have been investigated in a number of studies.
Ionization of atoms by a strong nonresonant laser field in the presence of a static field was first considered in Ref.~\cite{aru70}, in which the ionization rate was obtained with exponential accuracy (i.e., without preexponential factor), in the limits of tunneling and multiphoton ionization.
 The general exponential-accuracy solution, which includes the above
limits as special cases, was obtained in Ref.~\cite{ivl86}.
 Attempts to obtain a complete solution for electron detachment
from negative ions were made in Refs.~\cite{nik72,slo76,gao90}.
 However, the results obtained in Refs.~\cite{nik72,slo76,gao90} involve infinite series and are quite intricate.
 Moreover, Ref.~\cite{slo76} is concerned with the relatively simple,
but practically not very interesting, case (a circularly polarized
light perpendicular to the static field).
The analysis in Ref.~\cite{gao90} was limited to several simple cases where only one term of an infinite series was taken into account.
 There are also direct numerical solutions of the problem \cite{mer91}.
However these laborious calculations cannot provide the dependence of the ionization rate on the parameters for the entire parameter space.
Thus, at present there does not exist a comprehensive analysis of multiphoton detachment in the presence of a laser (or, more generally, ac) field for the whole range of the parameters of the problem

There has also been recent research on one- \cite{bry87,du88,man00} or few-photon \cite{man80,ost91,fro01} ionization in the presence of a static field.
 In particular, a ripplelike structure in the photodetachment cross
section was observed \cite{bry87}, which was explained by
interference of electron waves reflected from the barrier due to the
static field.

In this paper, we consider photodetachment from atoms or negative ions produced by a static and a nonresonant ac fields.
 We focus on the case of a linearly polarized ac field parallel to the static field, but the present treatment can be extended to the general case of elliptic polarization and arbitrary field orientations.
 We provide analytical results for the differential and total rates of detachment.
 As an application of the present theory, we discuss electron
emission from metals and semiconductors and from photosynthetic bio-complexes in the presence of a static and an ac fields.

The present theory is an extension of the Keldysh theory of ionization and electron detachment by a strong laser field \cite{kel64}.
The Keldysh theory provides the ionization and detachment rates and the energy spectrum of the emitted electrons in simple analytical forms.
The nonlinear effects of strong field are taken into account with the help of Volkov functions, which describe exactly the electron-field interaction in the final electron state.
The Volkov function for an electron in a harmonic field has, at least, two important advantages.
First, it has a simple analytical form that is convenient for calculations.
Second, the probability distribution of the ionization products in the Volkov-function basis directly provides the energy spectrum of the products.
This is due to the fact that in the limit of vanishing laser field the Volkov function tends to an energy eigenstate of a free particle, i.e., a plane wave.

The standard Volkov functions, obtained for the case of a harmonic field, cannot be used in the present case.
Hence an important step in the solution of the present problem is to find suitable functions providing the basis for the expansion of the final state of the electron.
An obvious candidate for such functions are the Volkov functions, describing an electron in a static and an ac fields \cite{nik72}.
These Volkov functions have a rather simple form.
However, they have the disadvantage that in the limit of the vanishing ac field, they do not reduce to eigenstates of the Hamiltonian for a particle in a uniform static field.

There is also an alternative basis, consisting of functions which have the important property that they become eigenstates of the above Hamiltonian in the limit of vanishing ac field \cite{gao90}.
These functions are expressed in terms of the Airy function and, as a result, unfortunately, are less convenient for calculations than Volkov functions.
Therefore, we use a combined approach, in which initially we calculate the final state of the electron in the basis of the Volkov functions, and then we transform this state into the second basis.
This approach yields simple analytical expressions for the rate of electron detachment and the energy spectrum of the emitted electrons.
These results allow us to obtain a comprehensive picture for the behavior of the detachment rate for a broad range of the parameters of the problem.
Our theory agrees with Keldysh theory \cite{kel64} in the limit of the vanishing static field.

The interaction of matter with electromagnetic fields can be described using different gauges.
Correspondingly, in analytical treatments of multiphoton and above-threshold ionization \cite{ebe91} various gauges have been used.
Most commonly, the so called velocity and length gauges have been used.
Although the exact solutions are gauge-independent, in approximate analytical theories, different gauges yield different results.
There are sound theoretical considerations \cite{rza04} as well as empirical evidence \cite{xie99,bau05} indicating that the length gauge has an advantage over other gauges.
In the present paper, following a number of previous studies, such  as, e.g., Ref.~\cite{kel64}, we use the length gauge.

In Sec. \ref{II} the general theory of bound-free transitions in a
time-dependent field is reviewed.
 In Sec. \ref{III} we introduce two sets of final-state wave
functions: the generalized Volkov functions and  functions that are a direct extension of the stationary states for a particle in a static uniform field.
 In Sec. \ref{IV} we consider the limit of a
very slowly varying field.
 Section \ref{V} presents a full solution of the problem.
 In contrast to previous studies, we use the
stationary-phase method, which provides a relatively simple
analytical solution.
 In Sec. \ref{VI} we describe several important
regimes of detachment, which follow from the above solution.
 Our solutions are much reacher than solutions in the presence of only one of the two fields, and a number of novel features are revealed.
 In Sec. \ref{VII} we discuss application
of the present theory to electron emission from solid surfaces.
 Section \ref{VIII} provides concluding remarks.
In Appendix \ref{A}, we obtain the expectation value of the energy in various states describing an electron affected simultaneously or separately by static and time-dependent fields.
In Appendix \ref{B}, we calculate a useful integral involving the Airy function.

\section{Quantum transitions due to a nonstationary field}
\label{II}

In this section we summarize our general approach, which is similar
to that developed by Keldysh,  Faisal, and Reiss
\cite{kel64,fai73,rei80} for a description of above-threshold
ionization \cite{ebe91}.
 The Hamiltonian of a quantum system (e.g.,
an atom or ion) in a field is
 \be
 H(t)=H_a+V(t),
 \label{1.1}\end{equation}
where $H_a$ is the atom Hamiltonian and $V(t)$ describes the
atom-field interaction, which induces transitions to the continuum
from the initial bound atomic state $|\phi_0\rangle$, an eigenstate
of $H_a$ with the energy $E_0$.
 We assume that only one electron interacts effectively with the
field.
 The field is assumed to be sufficiently weak, so that it does not
modify appreciably the initial state and its energy.
 Moreover, the field alternates periodically with the
frequency $\omega$ and is nonresonant.
 In view of the above assumptions, we can neglect transitions to other
discrete states and cast the wave function in the form,
 \be
|\Psi(t)\rangle=\beta(t)
|\phi_0\rangle+\sum_\lambda\beta_\lambda(t)|\psi_\lambda(t)\rangle,
 \e{9}
where the functions $|\psi_\lambda(t)\rangle$ form an orthonormal
basis of the continuum and $\beta(t)$ and $\beta_\lambda(t)$ are the
amplitudes of the initial and final states, respectively.

 For $t\gg2\pi/\omega$, the initial-level population is given by
 \be
 |\beta(t)|^2\approx e^{-Wt}.
\e{14}
 where $W$ is the rate of bound-free transitions.
The rate $W$ is obtained most easily by considering short times
$t\ll W^{-1}$, when Eq.~\rqn{14} becomes $|\beta(t)|^2\approx1-Wt$.
 Then the normalization condition for the wave function \rqn{9} implies
that
 \be
W=\lim_{t\rightarrow\infty}\frac{1}{t}\sum_\lambda
|\beta_\lambda(t)|^2.
 \e{23}

To obtain $\beta_\lambda(t)$ in Eq.~\rqn{23}, we use the time-dependent perturbation theory.
Namely, we note that for short times, in the first approximation,
 \be
\beta(t)\approx\exp\left(-\frac{i}{\hbar}E_0t\right).
 \e{2.3}
Assume that the functions $|\psi_\lambda(t)\rangle$ are the
solutions of the Schr\"{o}dinger equation
 \be
i\hbar\frac{\partial|\psi_\lambda\rangle}{\partial t}=
H_F(t)|\psi_\lambda\rangle.
 \e{2.2}
Here $H_F(t)$ is the Hamiltonian for the free (unbound) electron in
the field.
 Then one can show
with the help of \rqn{2.3} that \cite{fai73,rei80}
 \be
 \beta_\lambda(t)=-\frac{i}{\hbar}\int_0^tdt'\langle\psi_\lambda(t')
 |V(t')|\phi_0\rangle\exp\left(-\frac{i}{\hbar}E_0t'\right).
 \e{19}
In this paper, we use the common approximation \cite{kel64,fai73,rei80} in which the interaction of the detached electron with the field is taken into account, but the interaction with the atom is neglected.
Then
 \be
H_F(t)=\frac{\hat{\vec{P}}^2}{2m}+V(t),
 \e{2.5}
where $m$ is the electron mass and $\hat{\vec{P}}$ is the momentum
operator.

Note that the basis $\{|\psi_\lambda(t)\rangle\}$ satisfying
condition \rqn{2.2} is not unique, being defined with the accuracy to
an arbitrary time-independent unitary transformation.
 For a system with a time-independent Hamiltonian, there is a
preferable basis consisting of the eigenstates of the Hamiltonian.
 However, in the time-dependent case, such a preferable set of
solutions of the Schr\"{o}dinger equation is absent (the eigenstates
of the Hamiltonian are not solutions of the Schr\"{o}dinger
equation), and therefore more than one basis
$\{|\psi_\lambda(t)\rangle\}$ may appear convenient, depending on
the situation.
 In particular, in the present formalism, two types of states of the
electron in a field prove to be useful.
These states are described in the next section.

\section{Final-state wave functions}
\label{III}

Here the field-atom coupling is described in the dipole approximation and in the length gauge; correspondingly, the interaction term in Eqs.~\rqn{1.1} and \rqn{2.5} is
\be
 V(\vec{r},t)=-\vec{\cal F}(t)\cdot\vec{r}.
 \e{3.29}
Here $\vec{r}$ is the radius-vector of the electron initially bound to the atom, and $\vec{\cal F}(t)$ is the force exerted on the electron due to a uniform electric field $\vec{\cal F}(t)/e_0$, where $e_0$ is the (negative) electron charge, $e_0=-|e_0|$.

In this paper, the electric field $\vec{\cal F}(t)/e_0$ is assumed to be a superposition of a dc term $\vec{\cal E}/e_0$ and a harmonic field with amplitude $\vec F/e_0$,
\be
 \vec{\cal F}(t)=\vec{\cal E}+\vec F\sin\omega t.
 \e{3.29'}
 Below we use the units in which
\be
 \hbar=m=1.
 \e{3.3}

\subsection{Volkov functions}

\subsubsection{General formulas}

Equation \rqn{2.2} with the account of Eqs.~\rqn{2.5} and \rqn{3.29}, where $\vec{\cal F}(t)$ is an arbitrary function, has solutions in the form of plane waves,
\be
 \psi_{\vec{p}}(\vec{r},t)= e^{i\vec{P}(t)\cdot\vec{r}-
 i\int_0^td\tau P^2(\tau)/2},
 \e{3.5}
where
\be
 \vec{P}(t)=\vec{p}-\frac{e_0}{c}\vec{A}(t),
 \e{3.6}
Here $c$ is the speed of light and $\vec{A}(t)$ is defined by
\be
\dot{\vec{A}}=-c\vec{\cal F}(t)/e_0.
 \e{3.34}
It is not difficult to check that the function \rqn{3.5} indeed satisfies Eq.~\rqn{2.2}.
The functions of the form given in Eq.~\rqn{3.5} are called the Volkov function \cite{kel64,wol35}.

 The function $\psi_{\vec{p}}(\vec{r},t)$ is an eigenstate of the
momentum operator with the eigenvalue $\vec{P}(t)$ in Eq.~\rqn{3.6}.
 Note that in the length gauge assumed here, the vector potential
vanishes, and as a result the momentum coincides with the kinetic momentum \rqn{3.6}.
The vector $\vec{p}$ in Eq.~\rqn{3.6} is a constant of motion.

 The Volkov functions satisfy the orthonormality condition
 \be
\langle\psi_{\vec{p}}(t)|\psi_{\vec{p}'}(t)\rangle=
(2\pi)^3\delta(\vec{p}-\vec{p}')
 \e{3.18}
and the completeness condition
 \be
\int\frac{d\vec{p}}{(2\pi)^3}\psi_{\vec{p}}^*(\vec{r},t) \psi_{\vec{p}}(\vec{r}',t)=\delta(\vec{r}-\vec{r}').
 \e{3.30}
Therefore the Volkov functions for a given $\vec{\cal F}(t)$ form a basis of the Hilbert state of a free electron.

\subsubsection{Volkov functions for a sum of static and ac fields}

The general Volkov functions \rqn{3.5} are applicable, in particular, for the field given in Eq.~\rqn{3.29'} \cite{note1}.
Now in Eq.~\rqn{3.5}, $\vec{P}(t)$ is described by Eq.~\rqn{3.6} with
\be
\frac{e_0}{c}\vec{A}(t)=-\vec{\cal E}t+\frac{\vec{F}}{\omega}\cos\omega t.
 \e{3.31}

In the present paper, we focus on the important case, in which the ac electric field is linearly polarized in the direction of the constant electric field.
On choosing the latter direction as the $z$ axis, the interaction term \rqn{3.29} involved in Eqs.~\rqn{1.1} and \rqn{2.5} becomes
\be
 V(z,t)=-{\cal F}(t)z,
 \e{3.1}
where the force ${\cal F}(t)$, as implied by Eq.~\rqn{3.29'}, is given by
\be
 {\cal F}(t)={\cal E}+F\sin\omega t\ \ \ ({\cal E},F>0).
 \e{3.2}
Here, without loss of generality, we assume that ${\cal E}$ and $F$ are positive.
Now
\be
\vec{A}(t)=(0,0,A(t)),
 \e{3.33}
where
\be
\frac{e_0}{c}A(t)=-{\cal E}t+\frac{F}{\omega}\cos\omega t,
 \e{3.32}
and, correspondingly, Eq.~\rqn{3.6} becomes
 \be
 \vec{P}(t)=(p_x,p_y,P_z(t))
 \e{3.7}
with
\be P_z(t)=p_z+{\cal E}t-\frac{F}{\omega}\cos\omega t.
 \e{3.8}
Thus, the transverse momentum $\vec{q}=(p_x,p_y)$ is conserved,
whereas the $z$-component of the momentum varies with time, the
quantity $p_z$ being a constant of motion.

 Then Eq.~\rqn{3.5} becomes
\bea
&&\psi_{\vec{p}}(\vec{r},t)=\exp\left\{i\left[\vec{q}\cdot\vec{\rho}+
P_z(t)z-\left(\frac{p^2}{2}+U_p\right)t\right.\right.
\nonumber\\
&&-\frac{{\cal E}p_z}{2}t^2-\frac{{\cal E}^2}{6}t^3
+\frac{Fp_z}{\omega^2}\sin\omega t
-\frac{F^2}{8\omega^3}\sin 2\omega t\nonumber\\
&&\left.\left.-\frac{{\cal E}F}{\omega^2}t\sin\omega t +\frac{{\cal
E}F}{\omega^3}\cos\omega t\right]\right\},
 \ea{3.26}
where $\vec{\rho}=(x,y)$ and $U_p$ is the ponderomotive energy,
i.e., the average kinetic energy of the electron oscillating in the
harmonic field \cite{lan76},
 \be
U_p=\frac{F^2}{4\omega^2}=\frac{F^2}{4m\omega^2},
 \e{2.23}
the latter expression being written in the usual units.
In the derivation of Eq.~\rqn{3.26}, we omitted a constant phase.

The Volkov functions \rqn{3.26} have a relatively simple form and hence can be convenient for analytical calculations.
However, in the next subsection we show that the Volkov functions do not reduce to eigenstates of the energy operator in the limit of the vanishing ac field, and hence the Volkov functions are not suitable for obtaining the energy spectrum of the detached electrons.
Therefore, in Sec. \ref{IIIC} we introduce another set of solutions of Eq.~\rqn{2.2} with the account of Eqs.~\rqn{2.5}, \rqn{3.1}, and \rqn{3.2}, as a generalization of the stationary states for a particle in a uniform field.

\subsection{Static field: Stationary states and Volkov functions}
\label{IIIB}

In this subsection, we consider two types of the electron states for the case of a static, uniform field.

\subsubsection{Stationary states}

The stationary states of a particle in a static, uniform field have
the form \cite{lan77}
\be
\psi_{E\vec{q}}(\vec{r})=\psi_{E_z}(z)e^{i\vec{q}\cdot\vec{\rho}},
\e{3.9}
where $E$ is the energy of the state and
\be
 \psi_{E_z}(z)=\frac{2^{1/3}}{{\cal E}^{1/6}}{\rm Ai}(-u)
 \e{3.10}
is the energy eigenstate of one-dimensional motion of a particle in
a uniform field.
 Here
\be
 u=(2{\cal E})^{1/3}(z+E_z/{\cal E}),\ \ E_z=E-q^2/2
 \e{3.11}
 and
\be
 \mbox{Ai}(\xi)=\frac{1}{\pi}\int_0^\infty dv\cos\left(\frac{v^3}{3}+
 \xi v\right)
 \e{3.14}
is the Airy function \cite{abr64,note4}.
Note the asymptotic expressions for the Airy function \cite{abr64},
\bes{3.35}\bea
&&\mbox{Ai}(\xi)=\frac{1}{2\sqrt{\pi}\xi^{1/4}}\exp\left(-\frac{2}{3}\xi^{3/2}\right),\quad\ \xi\gg1\label{3.35a}\\
&&\mbox{Ai}(\xi)=\frac{1}{\sqrt{\pi}|\xi|^{1/4}}\sin\left(\frac{2}{3}|\xi|^{3/2}+\frac{\pi}{4}\right),\ \ -\xi\gg1,\quad\quad\quad
\ea{3.35b}\ese
whereas Ai$(0)\approx0.36$.
Thus, the Airy function decays exponentially for positive values of the argument and oscillates for negative values of the argument.

\subsubsection{Volkov functions for a static field}

Along with the standard basis of the stationary functions \rqn{3.9} described in textbooks, we consider an alternative basis of the static-field Volkov functions, following from Eq.~\rqn{3.26} for $F=0$,
 \bea
&\psi_{\vec{p}}(\vec{r},t)=&\exp\left\{i\left[\vec{q}\cdot\vec{\rho}+
(p_z+{\cal E}t)z\right.\right.
\nonumber\\
&&\left.\left.-\frac{p^2}{2}t-\frac{{\cal E}p_z}{2}t^2-\frac{{\cal
E}^2}{6}t^3 \right]\right\}.
 \ea{3.27}
 The two bases are related by a unitary transformation of the form
\be
\psi_{\vec{p}}(\vec{r},t)=\int_{-\infty}^\infty dE\,
h_q(p_z,E) \psi_{E\vec{q}}(\vec{r})e^{-iEt}.
 \e{3.12}
The function $h_q(p_z,E)$ here can be obtained as follows.
 Equation \rqn{3.12} has the form of a Fourier transform and hence can
be inverted:
\be
h_q(p_z,E)\psi_{E\vec{q}}(\vec{r})=\frac{1}{2\pi}
\int_{-\infty}^\infty dt\,\psi_{\vec{p}}(\vec{r},t)e^{iEt}.
 \e{3.13}
This integral can be transformed to the form proportional to
Eq.~\rqn{3.9} with the coefficient
\be
 h_q(p_z,E)=\frac{1}{\sqrt{{\cal E}}}\exp\left[\frac{ip_z}{\cal E}
\left(\frac{q^2}{2}+\frac{p_z^2}{6}-E\right)\right].
 \e{3.16}

It is easy to see that the function in Eq.~\rqn{3.16} satisfies the relations
 \bea
&&\int_{-\infty}^\infty\frac{dp_z}{2\pi}h_q^*(p_z,E')
h_q(p_z,E)=\delta(E-E'),\label{3.25}\\
&&\int_{-\infty}^\infty dE\,h_q(p_z,E)h_q^*(p_z',E)=2\pi\delta(p_z-p_z'),
 \ea{3.25'}
which ensure that the transformation in Eq.~\rqn{3.12} is unitary.
Using the unitarity of $h_q(p_z,E)$, we can invert the transformation in Eq.~\rqn{3.12}, as follows.
Multiplying the both sides of Eq.~\rqn{3.12} by $h_q^*(p_z,E')/(2\pi)$, integrating over $p_z$, and using Eq.~\rqn{3.25}, we obtain
\be
\psi_{E\vec{q}}(\vec{r})e^{-iEt}=\int_{-\infty}^\infty\frac{dp_z}{2\pi}\,h_q^*(p_z,E)\psi_{\vec{p}}(\vec{r},t).
 \e{3.28}

The Volkov function in Eq.~\rqn{3.27} has a relatively simple form, but it is not an eigenfunction of the energy operator.
In fact, the energy distribution in a Volkov state is infinitely broad.
Indeed, as follows from Eq.~\rqn{3.12}, this distribution is proportional to $|h_q(p_z,E)|^2$ and hence is uniform [cf.\ Eq.~\rqn{3.16}].
Note, however, that, as shown in Appendix \ref{A.1}, the expectation value of the energy in the state \rqn{3.27} is finite and equals $p^2/2$ [see Eq.~\rqn{A18}].

\subsection{Alternative basis of exact solutions for an electron in static and ac fields}
\label{IIIC}

As mentioned above, the Volkov functions are exact solutions of Eq.~\rqn{2.2}, and they constitute a basis for the Hilbert space of a free electron.
This basis is not unique---actually, any transformation of the Volkov functions with the help of a time-independent unitary operator produces a basis consisting of solutions of Eq.~\rqn{2.2}.
Here we introduce one such basis, consisting of functions which, in contrast to the Volkov functions, reduce to stationary states in the limit of the vanishing ac field.

In the case of a time-dependent field \rqn{3.2}, we consider a direct
extension of the unitary transformation \rqn{3.28},
 \be
 |\psi_{E\vec{q}}(t)\rangle=\int_{-\infty}^\infty
\frac{dp_z}{2\pi}h_q^*(p_z,E-U_p)|\psi_{\vec{p}}(t)\rangle,
 \e{3.19}
where $|\psi_{\vec{p}}(t)\rangle$ is the Volkov function for a particle  in the field \rqn{3.2} [see Eq.~\rqn{3.26}].
Equation \rqn{3.19} defines a new basis consisting of solutions of Eq.~\rqn{2.2} with the account of Eqs.~\rqn{2.5}, \rqn{3.1}, and \rqn{3.2}, in addition to the basis of the Volkov functions in Eq.~\rqn{3.26}.
Taking into account Eqs.~\rqn{3.18} and \rqn{3.25}, one can show that the functions \rqn{3.19} satisfy the normalization condition
\be
 \langle\psi_{E\vec{q}}(t)|\psi_{E'\vec{q}'}(t)\rangle=
(2\pi)^2\delta(\vec{q}-\vec{q}')\delta(E-E').
 \e{3.17}

In Eq.~\rqn{3.19}, the energy is shifted by the value of the ponderomotive energy, $U_p$.
This corresponds to the fact that the averaged motion of a particle affected simultaneously by a time-independent and a rapidly oscillating fields occurs as if in addition to the time-independent field there is another time-independent field with the potential energy equal to the ponderomotive energy \cite{lan76}.

To obtain the state $|\psi_{E\vec{q}}(t)\rangle$ in the coordinate representation, in Eq.~\rqn{3.19} we insert Eqs.~\rqn{3.16} and replace $|\psi_{\vec{p}}(t)\rangle$ by the function in Eq.~\rqn{3.26}; then the integration yields
\bea
 &\psi_{E\vec{q}}(\vec{r},t)=&\frac{2^{1/3}}{{\cal E}^{1/6}}
\exp\left\{i\left[\vec{q}\cdot\vec{\rho}+ \left(\frac{{\cal
E}}{\omega^2}-z\right)\frac{F}{\omega}
\cos\omega t\right.\right.\nonumber\\
&&\left.\left.-\frac{F^2}{8\omega^3}\sin 2\omega t-Et
\right]\right\}{\rm Ai}(-u(t)),
 \ea{3.21}
where
\be
 u(t)=(2{\cal E})^{1/3}\left(z+\frac{E_z-U_p}{\cal E}
+\frac{F}{\omega^2}\sin\omega t\right).
 \e{3.22}
As shown in Appendix \ref{A.2} [Eq.~\rqn{A26}], the mean energy in the state $|\psi_{E\vec{q}}(t)\rangle$ averaged over the oscillations of the electron is equal to $E$.

 One can invert the unitary transformation \rqn{3.19} with the help of
Eq.~\rqn{3.25'} to obtain
 \be
|\psi_{\vec{p}}(t)\rangle=\int_{-\infty}^\infty
dE\,h_q(p_z,E-U_p)|\psi_{E\vec{q}}(t)\rangle.
 \e{3.23}

\section{Ionization and detachment in the adiabatic limit}
\label{IV}

\subsection{Units of measurement}

Henceforth, we use dimensionless units \cite{lan77e} defined by relations \rqn{3.3} and by the equality
 \be
 E_0=-1/2.
 \e{4.1}
When the initial state is the ground state of a hydrogen atom, the present units coincide with the familiar atomic units where $\hbar=m=e_0^2=1$.
However, generally the present units differ from the atomic units.

The transition in any expression from the present dimensionless units to the conventional units is performed by the substitution
 \be
 X\rightarrow X/X_a
 \e{4.2}
for each quantity $X$ entering the expression, where $X_a$ is the unit of $X$ defined by Eqs.~\rqn{3.3} and \rqn{4.1}.
 For example, the units of frequency, momentum and force are
\bes{4.3}
\bea
&\omega_a=2|E_0|/\hbar,\label{4.3a}\\
&p_a=\sqrt{2m|E_0|},\label{4.3b}\\
&F_a=\omega_ap_a=\sqrt{m}(2|E_0|)^{3/2}/\hbar.
\ea{4.3c}
\ese

\subsection{Ionization in a static field}

Before developing a general theory, we consider the simple special
case of a field with a very low frequency (the adiabatic limit).
We start with ionization or detachment of an atom (or ion) by a static field ${\cal F}/e_0$.
In this case the ionization rate is given by
 \be
 W_{\rm st}=C({\cal F})e^{-2/(3|{\cal F}|)},
 \e{49}
where $C({\cal F})$ is the preexponential factor which depends on the specific details of the problem.

For instance, for field-induced ionization of a hydrogen atom in the ground state \cite{lan77a}
 \be
C({\cal F})=4/|{\cal F}|,
 \e{4.9}
whereas for electron detachment from an $s$-state in a potential well due to a short-range potential, such as, e.g., the potential in a negative ion, one has \cite{lan77e}
 \be
 C({\cal F})=\pi A^2|{\cal F}|.
 \e{4.4}
Here $A$ is the real coefficient in the asymptotic expression for the
electron wave function,
 \be
 \phi_0(r)=\frac{A}{r}e^{-r}\ \ (r\gg r_0),
 \e{43}
where $r_0$ is the radius of the potential ($r_0\alt 1$).
In particular, for the case of a weakly bound $s$-state \cite{lan77f}
 \be
 A=(2\pi)^{-1/2}\ \ (r_0\ll 1).
 \e{4.10}

\subsection{Static and ac fields: Adiabatic limit}
\label{IVC}

If the field $\cal F$ slowly varies with time, Eq.~\rqn{49} is
assumed to be still valid (the adiabatic limit).
In the case of the periodic field ${\cal F}(t)$ given in Eq.~\rqn{3.2}, the transition dynamics averaged over the field period is exponential with the rate obtained by averaging $W_{\rm st}$ over the period,
 \be
 W=\frac{1}{2\pi}\int_{-\pi}^\pi dv\,C({\cal E}+F\sin v)
 e^{-2/(3|{\cal E}+F\sin v|)}.
 \e{4.5}
This holds if the field frequency is much greater than the average transition rate,
 \be
 \omega\gg W.
 \e{4.12}

Let us calculate the integral in Eq.~\rqn{4.5}.
The exponential in the integrand of Eq.~\rqn{4.5} has the absolute maximum at $v=\pi/2$, where the field value is maximal, which occurs at times $(\pi/\omega)(2l+1/2)\ (l=0,1,\dots)$.
This maximum is the only local maximum for ${\cal E}>F$, whereas for ${\cal E}<F$, there is also a maximum at $v=-\pi/2$, which is
lower than the maximum at $v=\pi/2$.
 In the case considered here when ${\cal E}$ affects significantly
bound-free transitions, the contribution of the second maximum is
relatively small and hence will be neglected.
 Using the saddle-point method, we obtain
 \be
 W=\frac{1}{2}\sqrt{\frac{3}{\pi F}}({\cal E}+F)C({\cal E}+F)
 e^{-2/[3({\cal E}+F)]}.
 \e{4.6}
In particular, for photodetachment we obtain, with the account of Eq.~\rqn{4.4},
 \be
 W=\frac{A^2}{2}\sqrt{\frac{3\pi}{F}}({\cal E}+F)^2
 e^{-2/[3({\cal E}+F)]}.
 \e{4.7}

For validity of Eq.~\rqn{4.6}, the width of the peak in the
integrand of Eq.~\rqn{4.5}, given by the inverse square root of
the magnitude of the second derivative of the exponent
in \rqn{4.5} at $v=\pi/2$, should be much less than one.
This holds under the following conditions:
 \bea
&&{\cal E},F\ll 1,\label{4.8a}\\
&&F\gg{\cal E}^2.
 \ea{4.8c}
 Inequalities (\ref{4.8a}) coincide with the validity condition
of \rqn{49}, which states that the external field should be much weaker than the intraatomic field.
 In this case the field-induced barrier is so broad that the transition rate is exponentially small.

An analysis shows that for $F\ll{\cal E}^2$, the rate in Eq.~\rqn{4.5} is practically independent of the ac field and is approximately given by Eq.~\rqn{49} with ${\cal F}={\cal E}$, whereas for
 \be
F\agt{\cal E}^2
 \e{4.11}
the rate $W$ significantly depends on the ac field.
Thus, the condition \rqn{4.8c} (obtained previously in \cite{aru70}), in essence, coincides with the region in which the ac field significantly affects the transition rate.
 Similarly, when ${\cal E}$ is smaller than $F$, the static field significantly affects the transition rate when [cf.\ Eq.~\rqn{4.7}]
 \be
{\cal E}\agt F^2.
 \e{3.24}

Below we will recover Eq.~\rqn{4.7} as a special case of our theory for sufficiently small $\omega$.

\section{Detachment by the sum of dc and ac fields}
\label{V}

Here we apply the general theory of Sec. \ref{II} to detachment of negative ions by the field \rqn{3.2}.
 Using the Volkov functions, the wave function \rqn{9} in the coordinate representation takes the form
 \be
\Psi(\vec{r},t)=\beta(t)\phi_0(\vec{r})+\int
\frac{d\vec{p}}{(2\pi)^3}\,\beta_{\vec{p}}(t)\psi_{\vec{p}}(\vec{r},t).
\e{5.42}
 The amplitude of the transition from the initial state to a
Volkov function follows from Eq.~\rqn{19} with the account of Eqs.~\rqn{3.1} and \rqn{4.1},
\be
 \beta_{\vec{p}}(t)=i\int_0^tdt'\int d\vec{r}\,\phi_0(\vec{r})
{\cal F}(t')z\psi_{\vec{p}}^*(\vec{r},t')e^{it'/2}.
 \e{5.1'}
On changing the order of integration and using Eqs.~\rqn{3.5}, \rqn{3.7}, and \rqn{3.8}, Eq.~\rqn{5.1'} becomes
\be
 \beta_{\vec{p}}(t)=i\int d\vec{r}\,\phi_0(\vec{r})ze^{-i\vec{q}\cdot\vec{\rho}}
\int_0^tdt'\,{\cal F}(t')e^{if(t')},
 \e{5.1}
where
 \be
f(t)=\frac{t}{2}+\frac{1}{2}\int_0^td\tau\,P_z^2(\tau)
-P_z(t)z+\frac{q^2t}{2}.
 \e{5.2}

\subsection{Evaluation of the time integral in Eq.~\rqn{5.1}}

\subsubsection{Saddle points}
\label{VA}

We proceed to evaluate the time integral in Eq.~\rqn{5.1} by the
stationary-phase (or saddle-point) method.
The saddle points are the values of $t$ which satisfy the equation
\be \dot{f}(t)\equiv\frac{1}{2}+\frac{P_z^2(t)}{2} -{\cal
F}(t)z+\frac{q^2}{2}=0.
 \e{5.4}
Here we used Eq.~\rqn{5.2} and the relation, following from Eqs.~\rqn{3.6} and \rqn{3.34},
 \be
\dot{P}_z(t)={\cal F}(t).
 \e{5.17}
Equation \rqn{5.17} has the form of Newton's second law; this fact is a consequence of the Ehrenfest theorem.

Equation \rqn{5.4} is rather complicated.
Fortunately, as shown below, the coordinate $z$ and the momentum $\vec{q}$ are small in Eq.~\rqn{5.4}, and hence in the zero-order approximation we can neglect the last two terms in the right-hand side of Eq.~\rqn{5.4}, yielding $P_z^2(t)=-1$ or, in view of Eq.~\rqn{3.8},
\be
P_z(t)\equiv p_z+{\cal E}t-\frac{F}{\omega}\cos\omega t=is,
 \e{5.5}
where $s=\pm 1$.
Equation \rqn{5.5} implies that the saddle points are complex.

To solve Eq.~\rqn{5.5}, we write $t$ as the sum of real and imaginary parts, $t=t_r+it_i$, and separate the complex Eq.~\rqn{5.5} into two real equations
\bes{5.40}
\bea
&&p_z+{\cal E}t_r-\frac{F}{\omega}\cos\omega t_r\cosh\omega t_i=0,\label{5.40a}\\
&&{\cal E}t_i+\frac{F}{\omega}\sin\omega t_r\sinh\omega t_i=s,
 \ea{5.40b}
\ese
We assume that the quantity $p_z+{\cal E}t_r$ in Eq.~\rqn{5.40a} is small (the validity conditions for this assumptions are obtained below); hence it can be neglected in the zero-order approximation.
As a result, Eq.~\rqn{5.40a} yields the solutions $t_r=\tau_k$, where
 \be
\tau_k=\frac{\pi}{\omega}\left(k+\frac{1}{2}\right)\quad
(k=0,1,\dots)
 \e{5.7}
are the moments of the field extrema.
Inserting $t_r=\tau_k$ into Eq.~\rqn{5.40b} yields $t_i=\eta$, where $\eta$ is the solution of the equation
\be
\eta=\frac{s}{\cal E}-(-1)^k\frac{F}{\omega\cal E}\sinh\omega\eta.
\e{5.9}
Thus, we obtain that in the zero-order approximation the saddle points $t_k$ are given by
\be
t_k^{(0)}=\tau_k+i\eta.
\e{5.8}

As follows from Eq.~\rqn{5.4}, in the zero-order approximation
\be
\ddot{f}^{(0)}(t_k^{(0)})=P_z(t_k^{(0)})
\dot{P}_z(t_k^{(0)})=is{\cal F}(t_k^{(0)})=isb_k.
\e{5.10}
Here in the second equality we used Eqs.~\rqn{5.17} and \rqn{5.5}, whereas, in view of Eqs.~\rqn{3.2}, \rqn{5.8}, and \rqn{5.7},
\be
b_k\equiv{\cal F}(t_k^{(0)})={\cal E}+(-1)^kC_1,
\e{5.11}
where
\be
C_1=F\cosh\omega\eta=\sqrt{F^2+\omega^2(s-{\cal E}\eta)^2}.
\e{5.18}
The second equality in Eq.~\rqn{5.18} is implied by Eq.~\rqn{5.9}.

The quantity $p_z+{\cal E}t_r$, neglected in Eq.~\rqn{5.40a}, equals to a first approximation
\be
p_k=p_z+{\cal E}\tau_k.
 \e{5.6}
Equations \rqn{5.5} and \rqn{5.7} imply that $p_k=P_z(\tau_k)$, i.e., the quantity $p_k$ is the value of the $z$-component of the momentum at the moment $\tau_k$ for an electron detached in the vicinity of the $k$th saddle point.
The quantities $z,\ q$, and $p_k$ are the small parameters of the expansions below.

To obtain corrections to Eq.~\rqn{5.8}, we write the roots of Eq.~\rqn{5.4} in the form
\be
t_k=t_k^{(0)}+t_k^{(1)}+\dots,
\e{5.12}
where $t_k^{(n)}$ is the $n$th-order contribution into $t_k$.
Then we replace $t$ by Eq.~\rqn{5.12} in Eq.~\rqn{5.4} with the account of Eqs.~\rqn{3.2} and \rqn{3.8}, expand the trigonometric functions into the Taylor series in a neighborhood of $t_k^{(0)}$, and equate each sum of the terms of the same order to zero.
This yields $t_k^{(n)}$.
In particular,
\bes{5.13}
\bea
&&t_k^{(1)}=-isz-p_k/b_k,\label{5.13a}\\
&&t_k^{(2)}=i\frac{z^2}{2}\left(\frac{S_1}{b_k}-sb_k\right)
-i\frac{S_1p_k^2}{2b_k^3}+i\frac{sq^2}{2b_k},
\ea{5.13b}
\ese
where
\be
S_1=(-1)^kF\omega\sinh\omega\eta=\omega^2(s-{\cal E}\eta),
\e{5.16}
the latter equality following from Eq.~\rqn{5.9}.

Furthermore, we calculate $f(t_k)$.
To this end, we note that the explicit form for $f(t)$ in Eq.~\rqn{5.2} is given by the expression in the square brackets in Eq.~\rqn{3.26}, where $\vec{q}\cdot\vec{\rho}$ should be replaced by $t/2$.
(In this expression a constant term is neglected, which provides an unimportant constant phase.)
Inserting in this expression Eq.~\rqn{5.12} instead of $t$ and performing the Taylor expansion of the resulting expression about $t_k^{(0)}$, we obtain the expansion of $f(t_k)$,
\be
f(t_k)=f^{(0)}(t_k)+f^{(1)}(t_k)+\dots,
\e{5.14}
where $f^{(n)}(t_k)$ is the contribution of the $n$th order of magnitude.
In particular,
\bes{5.15}
 \be
 f^{(0)}(t_k)=\frac{{\cal E}^2\tau_k^3}{6}
+\left(U_p+\frac{1}{2}\right)\tau_k+iC_0,
 \e{5.15a}
\be
f^{(1)}(t_k)=-isz-\left[\frac{{\cal E}}{2}(\tau_k^2+\eta^2)
+(-1)^k\frac{C_1}{\omega^2}\right]p_k,
\e{5.15b}
\be
f^{(2)}(t_k)=isb_k\frac{z^2}{2}+\left(\tau_k+ic_k\right)
\frac{p_k^2}{2}+(\tau_k+i\eta)\frac{q^2}{2},
\e{5.15c}
\ese
where
\be
C_0=\left(U_p+\frac{1}{2}\right)\eta-\frac{{\cal E}^2\eta^3}{6}
-(-1)^kC_1\frac{s+3{\cal E}\eta}{4\omega^2}
+{\cal E}\frac{s-{\cal E}\eta}{\omega^2},
\e{5.50}
\be
c_k=\eta-s/b_k.
\e{5.30}

The above expansions \rqn{5.12} and \rqn{5.14} are valid, if the
quantities $z$, $q$, and $p_k$ are sufficiently small. As implied by
the imaginary part of Eq.~\rqn{5.15c} [cf.\ the last integral in
\rqn{5.1}], the coordinate and momentum of the electron at the moment
$\tau_k$ have the bounds
 \be
|z|\alt(sb_k)^{-1/2},\ \ |p_k|\alt c_k^{-1/2},\ \
|q|\alt\eta^{-1/2}.
 \e{5.32}
Thus, the condition that $q$, $p_k$, and $z$ have bounded
distributions (which is a weaker requirement than the smallness of
$z$, $q$, and $p_k$) is equivalent to the simultaneous inequalities
 \be
{\rm (a)}\ sb_k>0,\ \ {\rm (b)}\ \eta>s/b_k,
 \e{5.20}
and $\eta>0$.
Condition (a) in Eq.~\rqn{5.20} ensures that the integrand of the time integral in \rqn{5.1} has a maximum at $t_k$ [cf.\ Eq.~\rqn{5.10}], implying that all roots of Eq.~\rqn{5.4} are saddle points, whereas the condition $\eta>0$, which is a consequence of Eq.~\rqn{5.20}, requires that all saddle points are located in the upper half space of the complex $t$ plane [cf.\ Eq.~\rqn{5.8}].

The conditions in Eq.~\rqn{5.20} imply the value of $s$, as follows.
Inequality (b) in Eq.~\rqn{5.20} can be recast as $\eta>1/(sb_k)$, since $s^2=1$, or $sb_k\eta>1$, in view of condition (a) in Eq.~\rqn{5.20}.
Hence, we can write
\bea
&&sb_k\eta-1=s{\cal E}\eta-1+(-1)^ks\eta F\cosh\omega\eta\nonumber\\
&&=(-1)^ks\eta F\cosh\omega\eta\left(1
-\frac{\tanh\omega\eta}{\omega\eta}\right)>0,
\ea{5.22}
where we used Eqs.~\rqn{5.11} and \rqn{5.18} in the first equality and Eq.~\rqn{5.9} in the second equality.
Since the last factor in the parentheses in Eq.~\rqn{5.22} is positive and $\eta>0$, the inequality in Eq.~\rqn{5.22} yields
\be
s=(-1)^k.
\e{5.23}

For the moments $\tau_k$ with even (odd) $k$, the ac field is directed along (opposite to) the static field.
Correspondingly, the cases with even and odd $k$ play different roles in the process of detachment.
For even $k$, Eq.~\rqn{5.11} implies that $b_k>0$, and condition (a) in Eq.~\rqn{5.20} always holds.
However, for odd $k$, when $F<{\cal E}$ and $\omega$ is sufficiently low, condition (a) in Eq.~\rqn{5.20} is violated, which means that there are no saddle points in such cases.
[Indeed, then Eq.~\rqn{5.11} implies that $b_k>0$, whereas $s=-1$, in view of Eq.~\rqn{5.23}, yielding a violation of condition (a) in Eq.~\rqn{5.20}.]
This fact can be easily understood, if one takes into account that for $F<{\cal E}$ the moments $\tau_k$ with odd $k$ correspond to the {\em field minima}, and hence such moments cannot contribute appreciably to detachment, at least, in the adiabatic (very low $\omega$) limit (cf.\ Sec.~\ref{IVC}).
Though the case of odd $k$ with $F>{\cal E}$ involves local maxima of the field magnitude equal to $F-{\cal E}$, these maxima are less than the maxima equal to $F+{\cal E}$, obtained for even $k$.
Hence, the case of odd $k$ produces a negligible contribution to the detachment rate, unless ${\cal E}$ is very small \cite{note5}.

Therefore, below we restrict the consideration to the saddle points
corresponding to the field global maximum, i.e., to the case of
$k=2l\ (l=0,1,\dots)$.
 Then $s=1$ [see Eq.~\rqn{5.23}], and the time integral in Eq.~\rqn{5.1} is, in view of Eq.~\rqn{5.10},
 \bea
 &&\int_0^tdt'{\cal F}(t')e^{if(t')}\approx\sum_{l=0}^N{\cal F}(t_{2l}^{(0)})e^{if(t_{2l})}\nonumber\\
 &&\times\int_{-\infty}^\infty dt'\,e^{-b_0(t'-t_{2l})^2/2}=\sqrt{2\pi b_0}\sum_{l=0}^Ne^{if(t_{2l})},\quad
 \ea{5.34}
where $N=[\omega t/2\pi-1/4]$, $[\dots]$ denotes the integer part,
and
 \be
b_0=b_{2l}={\cal E}+C_1,\ \
 C_1=\sqrt{F^2+\omega^2(1-{\cal E}\eta)^2}.
 \e{2.36}
Note that now in Eq.~\rqn{5.30} $c_k=c_{2l}=c_0$, where, in view of \rqn{5.23},
\be
c_0=\eta-1/b_0.
\e{5.78}

\subsubsection{Asymptotic expansions for some parameters}

The present theory involves a number of auxiliary parameters, which enter the analytical results.
It is useful to have simple formulas, providing approximations or estimates of the parameters of the present theory.
Here we provide asymptotic expansions for some parameters.

We begin with $\eta$, which is a solution of the transcendental equation \rqn{5.9}.
In the case of even $k$, Eq.~\rqn{5.9}, in view of Eq.~\rqn{5.23}, becomes
\be
{\cal E}\eta=1-\frac{F}{\omega}\sinh\omega\eta.
\e{5.81}
This equation has a unique solution, which decreases with $\omega$.
It is possible to obtain asymptotic expansions for $\eta$ in different limits.

In particular, for small $\omega$, one can expand $\sinh\omega\eta$ in Eq.~\rqn{5.81} in powers of $\omega\eta$, which allows us to obtain the expansion,
\be
\eta=\frac{1}{{\cal E}+F}-\frac{F\omega^2}{6({\cal E}+F)^4}
-\frac{F({\cal E}-9F)\omega^4}{120({\cal E}+F)^7}+O(\omega^6).
\e{5.24}
This series is rapidly converging, and hence can be truncated at any term to a good approximation, when
\be
\omega\ll{\cal E}+F.
\e{5.26}
Note that this condition is equivalent to the requirement that, at least, one of the following two inequalities hold, either $\omega\ll{\cal E}$ or $\omega\ll F$.

In the opposite case of large values of $\omega$, in the zero-order approximation, the term on the left-hand side of Eq.~\rqn{5.81} can be neglected, yielding $\eta^{(0)}=\omega^{-1}{\rm arcsinh}(2\omega/F)$.
Then, on expanding $\sinh\omega\eta$ in Eq.~\rqn{5.81} in a Taylor series in the vicinity of $\eta^{(0)}$ and taking into account that \cite{abr64}
 \be
 {\rm arcsinh}\,x=\ln 2x-1/(4x^2)+O(x^{-4}),
 \e{5.72}
we obtain the expansion of $\eta$ up to second order,
\be
 \eta\approx\frac{1}{\omega}\left\{
\left[1-\frac{{\cal E}}{\omega}+ \frac{{\cal E}^2}{\omega^2}
\left(1-\frac{1}{2}\ln\frac{2\omega}{F}\right)\right]
\ln\frac{2\omega}{F}-\frac{F^2}{4\omega^2}\right\}.
 \e{5.28}
This series is rapidly converging when
\be
\omega\gg{\cal E}\ln\frac{\omega}{F},\,F.
\e{5.69}

All other parameters entering the present formulas are elementary functions of $\eta$, and hence asymptotic expressions for them can be easily obtained using Eqs.~\rqn{5.24} and \rqn{5.28}.
In particular, below we show asymptotic expressions for $b_0$ and $c_0$ defined in Eqs.~\rqn{2.36} and \rqn{5.78}.
It appears that $b_0$ increases with $\omega$,
\bes{5.29}
\be
b_0\approx{\cal E}+F+\frac{F\omega^2}{2({\cal E}+F)^2},\ \
\omega\ll{\cal E}+F,
\e{5.29a}
\be
b_0\approx\omega-{\cal E}\ln\frac{2\omega}{eF},\ \
\omega\gg{\cal E}\ln\frac{\omega}{F},\,F;
\e{5.29b}
\ese
whereas $c_0$ as a function of $\omega$ increases for small $\omega$
and decreases for large $\omega$,
\bes{5.31}
\bea
&&c_0\approx\frac{F\omega^2}{3({\cal E}+F)^4},\ \
\omega\ll{\cal E}+F,\label{5.31a}\\
&&c_0\approx\frac{1}{\omega}\ln\frac{2\omega}{eF},\ \
\omega\gg{\cal E}\ln\frac{\omega}{F},\,F.
\ea{5.31b}
\ese
Note that the asymptotic expansions for $C_1$ are easily obtained from Eqs.~\rqn{5.29}, since, according to Eq.~\rqn{2.36}, $C_1=b_0-\cal E$.

Below we will need a rough estimate of $b_0$ for the whole range of the values of the input parameters ${\cal E},\ F$, and $\omega$.
The asymptotic expansions in Eqs.~\rqn{5.29} imply such an estimate in the form
\be
b_0\sim{\cal E}+F+\omega.
\e{5.84}

\subsubsection{Validity conditions for Eq. \protect\rqn{5.34};
higher-order corrections}

Consider in detail the assumptions made in Sec. \ref{VA}.
The above use of the stationary-phase method holds if the effective
range of time integration in Eq.~\rqn{5.34}, which is on the order of
$b_0^{-1/2}$, is much less than $t_{k+1}-t_k=\pi/\omega$.
In view of Eqs.~\rqn{5.29}, this requirement is satisfied under the
condition
\be
\omega\ll 1.
\e{5.25}
In the usual units, this conditions means that $\omega\ll|E_0|/\hbar$, i.e., the field frequency should be much less than the ionization potential divided by $\hbar$.

Furthermore, in Sec. \ref{VA} we expanded the trigonometric functions entering Eqs.~\rqn{5.2} and \rqn{5.4} in the small parameters, which is allowed if
\be
\omega|t_k^{(m)}|\ll 1\quad(m\ge 1).
\e{5.33}
Using Eqs.~\rqn{5.13} and the limits for the values of the quantities in Eq.~\rqn{5.32} with the account of Eqs.~\rqn{5.24}, \rqn{5.28}, \rqn{5.29}, and \rqn{5.31}, it can be shown that the inequalities \rqn{5.33} hold under the conditions \rqn{4.8a}, \rqn{4.8c}, and \rqn{5.25}.

To check the convergence of the series \rqn{5.12} and \rqn{5.14} for $k=2l\ (l=0,1,\dots)$, we considered the asymptotic formulas for Eqs.~\rqn{5.13} and \rqn{5.15} in the limits of low and high $\omega$.
We also obtained higher-order terms, up to fifth order, in expansions \rqn{5.12} and \rqn{5.14} with the help of the software package Mathematica \cite{wol96}.
The latter expressions are rather unwieldy and are not shown here.
The above analysis shows that
under conditions \rqn{4.8a}, \rqn{4.8c}, and \rqn{5.25}, the terms in the series \rqn{5.12} and \rqn{5.14} involving only $z$ and $q$ decrease rapidly with their order and, starting from third order, such terms can be neglected in \rqn{5.34}.

In contrast, terms involving $p_k$ require a special consideration.
Indeed, Eqs.~\rqn{5.32} and \rqn{5.31a} imply that
\be
|p_{2l}|\alt({\cal E}+F)^2/(\omega\sqrt{F})\quad
(\omega\ll{\cal E}+F),
\e{5.35}
i.e., the width of the distribution for $p_{2l}$ increases without bound with decreasing $\omega$, thus violating the above assumption of the smallness of $p_{2l}$.
However, for fixed $\omega$, the terms proportional to a power of
$p_{2l}$ decrease with the power.
Correspondingly, our estimates of the terms in the expansion \rqn{5.14} show that when $\omega$ is not too small,
\be
 \omega^2\gg({\cal E}+F)^4/F,
 \e{5.38}
some third-order terms yield appreciable contributions and should be kept in the expansion, whereas the other third- and higher-order terms can be neglected.
In particular, we keep in the expansion \rqn{5.14} (for $k=2l$) the third-order terms involving $p_{2l}$,
except for the term $\sim z^2p_{2l}$, which proves to be much less than the term $\sim z^2$, so that we obtain
\bes{5.37}
\be
f^{(3)}_{p}(t_{2l})=-C_2p_{2l}^3/3-q^2p_{2l}/(2b_0),
\e{5.37a}
where
\be
C_2=\frac{{\cal E}^2(1+\omega^2\eta^2)+2{\cal E}C_1
-3{\cal E}\omega^2\eta+F^2+2\omega^2}{2b_0^3}.
\e{5.37b}
\ese
Note that the frequency interval in Eq.~\rqn{5.38} coincides with the region, where the detachment rate depends significantly on $\omega$ [see the remark after Eq.~\rqn{6.11}].

\subsection{Detachment rate and electron spectrum}
\label{VC}

\subsubsection{Momentum (Volkov-function) representation}

Inserting Eq.~\rqn{5.34} into \rqn{5.1}, we obtain
\be
\beta_{\vec{p}}(t)=i\sqrt{2\pi b_0}\,I
\sum_{l=0}^Ne^{if_{p}(t_{2l})},
\e{5.39}
where $f_{p}(t_{2l})=f(t_{2l})|_{z=0}$ and
\be
I=\int d\vec{r}\,\phi_0(\vec{r})z
e^{-i\vec{q}\cdot\vec{\rho}+z-b_0z^2/2}.
 \e{5.44}
An analysis shows that the interval $-\infty<z<\zeta_0$ provides a negligibly small contribution into the integral over $z$ in Eq.~\rqn{5.44}, where $\zeta_0$ is a number satisfying $1\ll\zeta_0\ll b_0^{-1/2}$.
 (This result will be shown to be independent of the specific value of $\zeta_0$.)
In other words, $I$ is determined by large positive values of $z$, which reflects the fact that the electron tunnels in the positive
direction of the $z$ axis when the force is positive, ${\cal
F}(\tau_{2l})={\cal E}+F>0$.

Moreover, we assume, for simplicity, that the initial state is an $s$-state.
This allows us to use the asymptotics \rqn{43}, which now can be further simplified,
\be
\phi_0(r)\approx(A/z)\exp[-z-\rho^2/(2z)]\quad(z\gg1).
 \e{5.85}
Inserting Eq.~\rqn{5.85} into Eq.~\rqn{5.44} yields
\bea
&I&=A\int_{\zeta_0}^\infty dz\,e^{-b_0z^2/2}
\int d\vec{\rho}\,e^{-i\vec{q}\cdot\vec{\rho}-\rho^2/(2z)}\nonumber\\
&&=2\pi A\int_{\zeta_0}^\infty dz\,ze^{-b_0z^2/2-q^2z/2}\nonumber\\
&&=\frac{2\pi A}{b_0}f_1\left(\frac{q^2}{\sqrt{8b_0}}\right).
 \ea{5.52}
Here
\be
f_1(\xi)=1-\sqrt{\pi}\xi e^{\xi^2}{\rm erfc}(\xi),
 \e{5.87}
where ${\rm erfc}(\xi)$ is the error function \cite{abr64}.
To obtain the last equality in Eq.~\rqn{5.52}, we replaced $\zeta_0$ by 0, which practically does not change the result.

Equation \rqn{5.52} can be simplified as follows.
As shown below, in Sec.~\ref{VB3}, the range of the values of $q^2$ for detached electrons is given by $q^2\alt{\cal E}+F+\omega$.
Therefore, in Eq.~\rqn{5.52}, the quantity $q^2/\sqrt{8b_0}\alt({\cal E}+F+\omega)b_0^{-1/2}\sim({\cal E}+F+\omega)^{-1/2}\ll1$, where we took into account Eqs.~\rqn{5.84}, \rqn{5.25}, and \rqn{4.8a}.
Thus, we obtain that $q^2/\sqrt{8b_0}\ll1$.
Taking into account that $f_1(\xi)=1-\sqrt{\pi}\xi+2\xi^2+O(\xi^3)$, we can set to a first approximation $f_1(q^2/\sqrt{8b_0})\approx1$ in Eq.~\rqn{5.52}, yielding
\be
I\approx\frac{2\pi A}{b_0}.
 \e{5.52'}

The detachment rate \rqn{23} becomes now
\be
W=\lim_{t\rightarrow\infty}\frac{1}{t}\int\frac{d\vec{p}}{(2\pi)^3}
|\beta_{\vec{p}}(t)|^2,
 \e{5.67}
where $\beta_{\vec{p}}(t)$ is given by Eq.~\rqn{5.39}.
The Volkov-function representation of the final state does not provide the energy distribution of the detached electrons, which is directly measurable in experiments.
Moreover, this representation is not very convenient for calculations, since the summation in Eq.~\rqn{5.39} is difficult to perform analytically, due to the imaginary terms proportional to $\tau_{2l}^3$ and $\tau_{2l}^2$ in the exponent in \rqn{5.39} [see Eqs.~\rqn{5.15}].
 Therefore, it is advantageous to transform the final state of the electron from the Volkov-function representation to the representation in the basis \rqn{3.21}, as follows.

\subsubsection{Energy representation}
\label{VC2}

 On inserting the function in Eq.~\rqn{3.23} written in the coordinate representation into Eq.~\rqn{5.42}, the system wave function becomes
 \be
\Psi(\vec{r},t)=\beta(t)\phi_0(\vec{r})+\int_{-\infty}^\infty dE\int
\frac{d\vec{q}}{(2\pi)^2}\beta_{E\vec{q}}(t)
\psi_{E\vec{q}}(\vec{r},t).
\e{5.43}
Here
 \be
\beta_{E\vec{q}}(t)=\int_{-\infty}^\infty\frac{dp_z}{2\pi}
h_q(p_z,E-U_p)\beta_{\vec{p}}(t).
 \e{5.45}
On inserting here Eqs.~\rqn{3.16} and \rqn{5.39} with the account of Eq.~\rqn{5.52'} and changing the variable of integration $p_z\rightarrow p_{2l}$ by Eq.~\rqn{5.6}, the terms proportional to $\tau_{2l}^3$ and $\tau_{2l}^2$ cancel out and we obtain
 \bea
&\beta_{E\vec{q}}(t)&=iA\sqrt{\frac{2\pi}{b_0\cal E}} e^{-C_0-\eta
q^2/2}\sum_{l=0}^Ne^{i\Delta E\tau_{2l}}\nonumber\\
&&\times\int_{-\infty}^\infty dp_{2l}e^{iC_3p_{2l}^3/(6{\cal
E})-c_0p_{2l}^2/2-ia_qp_{2l}}.
\ea{5.46}
Here
 \be
\Delta E=E+1/2=E-E_0
 \e{5.47}
is the change of the average electron energy as a result of detachment and
 \be
C_3=1-2{\cal E}C_2,\ \ a_q=\frac{E}{\cal E}+z_1-\frac{d_1q^2}{2\cal
E},
 \e{5.48}
where
 \be
z_1=\frac{{\cal E}\eta^2}{2}+\frac{C_1}{\omega^2},\ \
d_1=\frac{C_1}{b_0}.
 \e{5.54}

Equation \rqn{5.46} can be further simplified.
We evaluate the integral in Eq.~\rqn{5.46} with the help of the change of variable
$p_{2l}\rightarrow p'=p_{2l}+i{\cal E}c_0/C_3$, which eliminates the quadratic term in the integrand; then we sum the geometric progression.
As a result, we obtain
 \bea
&\beta_{E\vec{q}}(t)=&iA\left(\frac{2{\cal E}}{C_3}\right)^{1/3}
\frac{(2\pi)^{3/2}}{\sqrt{b_0\cal E}}{\rm Ai}
\left(\frac{d_1q^2/2-E'}{\Delta_2}\right)\nonumber\\
&&\times e^{-C_5/2-E'/(2\Delta_1)-C_4q^2/2 +i\pi\Delta E/(2\omega)}
\nonumber\\
&&\times\frac{1-e^{2i\pi(N+1)\Delta E/\omega}} {1-e^{2i\pi\Delta
E/\omega}},
 \ea{5.49}
where
 \be
C_4=\eta-\frac{c_0d_1}{C_3},\quad C_5=2C_0+\frac{{\cal
E}^2c_0^3}{3C_3^2},
 \e{5.51}
 \be
\Delta_1=\frac{C_3}{2c_0},\quad
\Delta_2=\left(\frac{{\cal E}^2C_3}{2}\right)^{1/3},
 \e{5.61}
 \be
E'=E-E_c,\ \ E_c=U_p-{\cal E}z_0,\ \
 z_0=z_1-\frac{{\cal E}c_0^2}{2C_3}.
 \e{5.79}

The detachment rate \rqn{23} has now the form
 \be
W=\int_{-\infty}^\infty dE\int \frac{d\vec{q}}{(2\pi)^2}\,
\frac{\partial^2W}{\partial E\partial\vec{q}},
 \e{5.53}
where the differential rate
 \be
\frac{\partial^2W}{\partial E\partial\vec{q}}
=\lim_{t\rightarrow\infty}\frac{1}{t}|\beta_{E\vec{q}}(t)|^2.
 \e{5.80}
On inserting here Eq.~\rqn{5.49}, one can show that the rate of the detachment of an electron with given values of $E$ and the transverse momentum $\vec{q}$ is
\bes{5.55}
 \be
\frac{\partial^2W}{\partial E\partial\vec{q}}=\sum_{n=-\infty}^\infty
w_n(q)\delta(E-E_n),
 \e{5.55a}
where
 \bea
&w_n(q)=&\frac{2^{2/3}A^2\omega^2}{{\cal
E}^{1/3}b_0C_3^{2/3}} e^{-C_5-E_n'/\Delta_1-C_4q^2}\nonumber\\
&&\times{\rm Ai}^2\left(\frac{d_1q^2/2-E_n'}{\Delta_2}\right),
 \ea{5.55b}
 \ese
 \bea
&&E_n=n\omega-1/2=n\hbar\omega+E_0,\label{5.59}\\
&&E_n'=E_n-E_c=E_n-U_p+{\cal E}z_0.
 \ea{5.59'}
The last expression in Eq.~\rqn{5.59} is written in the usual units.

The detachment rate of electrons with a given $E$, irrespective of
$\vec{q}$, is obtained by integration over $\vec{q}$ in Eq.
\rqn{5.53} with the account of Eq.~\rqn{5.55},
\bes{5.62}
 \be
\frac{dW}{dE}=\sum_{n=-\infty}^\infty W_n\delta(E-E_n),
 \e{5.62a}
where
 \be
W_n=\frac{2^{2/3}\pi A^2\omega^2B_n}{{\cal E}^{1/3}b_0C_3^{2/3}C_4}
e^{-C_5-E_n'/\Delta_1}.
 \e{5.62b}
\ese
Here
 \be
B_n=\int_0^\infty{\rm
Ai}^2\left(\frac{d_2u-E_n'}{\Delta_2}\right)e^{-u}\,du,
 \e{5.64}
where
 \be
d_2=d_1/(2C_4).
 \e{5.65}

The total detachment probability is the integral of Eq.~\rqn{5.62a} over the energy,
 \be
W=\sum_{n=-\infty}^\infty W_n.
 \e{5.66}
 Equations \rqn{5.49}, \rqn{5.55}, \rqn{5.62}, and \rqn{5.66} present the main results of the present paper.

\subsubsection{Asymptotic expressions}

In Sec.~\ref{VC2} we introduced a number of new parameters.
To get an idea about the values of these parameters, simple asymptotic expressions for them can be obtained with the help of Eqs.~\rqn{5.24}, \rqn{5.28}, \rqn{5.29}, and \rqn{5.31}.
As an example, we show for some of the above parameters the following simplified expressions valid in the limits of low and high frequencies,
 \bea
d_2\approx
\left\{\begin{array}{ll}F/2,&\ \omega\ll{\cal E}+F\\
\omega/2,&\ \omega\gg{\cal E}\ln(\omega/F),\,F
\end{array}\right.
 \ea{5.70}
 \be
C_3\approx \left\{\begin{array}{ll}F/({\cal E}+F),
&\ \omega\ll{\cal E}+F\\
1,&\ \omega\gg{\cal E}\ln(\omega/F),\,F
\end{array}\right.
 \e{5.71}
\bes{6.14}
 \be
z_0\approx\frac{F}{\omega^2}+\frac{1}{2({\cal E}+F)}
-\frac{F\omega^2}{8({\cal E}+F)^4},\ \ \omega\ll{\cal E}+F,
 \e{6.14a}
 \be
z_0\approx\frac{1}{\omega} -\frac{{\cal E}}{2\omega^2},\ \
\omega\gg{\cal E}\ln\frac{\omega}{F},\,F.
 \e{6.14b}
 \ese
 \be
C_4\approx \left\{\begin{array}{ll}({\cal E}+F)^{-1},
&\ \omega\ll{\cal E}+F\\
\omega^{-1},&\ \omega\gg{\cal E}\ln(\omega/F),\,F
\end{array}\right.
 \e{5.86}

\subsubsection{Discussion}
\label{VB3}

Let us discuss the above results.
Equation \rqn{5.55a} or \rqn{5.62a} describes the spectrum of the detached electrons.
These equations show that the energy distribution of the detached
electrons consists of narrow equidistant peaks separated by
$\hbar\omega$.
The energy $E_n$ of the electrons in a given peak corresponds to absorption of $n$ photons, if $n>0$, or emission of $-n$ photons, if $n<0$ [cf.\ Eq.~\rqn{5.59}].
 The $n$th term in Eq.~\rqn{5.55a} or \rqn{5.62a} provides the rate of increase per one atom of, respectively, the transverse-moment distribution or the total number of detached electrons with the energy $E_n$.

The distribution of the electrons over the transverse momentum $\vec{q}$ is axially symmetric.
According to Eq.~\rqn{5.55b}, the $q$ distribution for a given energy is given by the Gaussian factor $e^{-C_4q^2}$, which is independent of $n$, multiplied by the last factor in Eq.~\rqn{5.55b}, which is $n$-dependent.
The factor $e^{-C_4q^2}$ shows that with the probability close to one we have $q^2\alt C_4^{-1}\sim{\cal E}+F+\omega$, the latter relation being implied by Eq.~\rqn{5.86}.
The asymptotic expressions for the Airy function \rqn{3.35} imply that the last factor in Eq.~\rqn{5.55b} can limit further the above interval for $q^2$.
Moreover, for positive $E_n'$, this factor can provide {\em oscillations} of the number of the detached electrons as a function of $q$ [cf.\ Eq.~\rqn{3.35b}].

\begin{figure}[htb]
\begin{center}
\includegraphics[width=8cm]{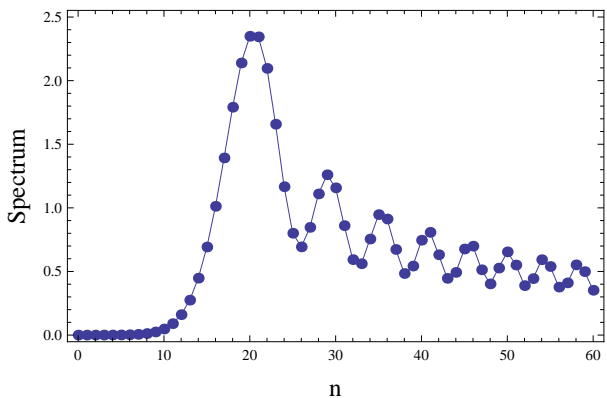}
\end{center}
\caption{The spectrum, $W_n$, in units of $2\times10^{-9}|E_0|/\hbar$ is shown by dots.
Lines joining the dots are a guide for eyes. Here ${\cal E}=0.013,\ F=0.05$, and $\omega=0.01$ in the dimensionless units.
}
 \label{f1}\end{figure}

The total energy spectrum integrated over $q$, Eqs.~\rqn{5.62}, also generally demonstrates oscillations resulting due to the static field, see Fig.~\ref{f1}.

\section{Special cases}
\label{VI}

\subsection{The total rate with exponential accuracy}

As a preparation for the analysis of the above general formulas, we provide here the asymptotic expressions for the quantity $e^{-2C_0}$,
where $C_0$ is given in Eq.~\rqn{5.50} with $k=0$ [and hence $s=1$, as implied by Eq.~\rqn{5.23}].
In view of Eqs.~\rqn{5.39} and \rqn{5.15a}, the quantity $e^{-2C_0}$ provides the zero-order approximation to $W$,
 \be
W\propto e^{-2C_0}.
 \e{6.1}
This approximation is known as an approximation with ``exponential accuracy'', since it involves a neglect of higher-order terms in the exponent in Eq.~\rqn{6.1}.
It is interesting that Eq.~\rqn{6.1} coincides with the formula for the rate of transmission through a one-dimensional triangular barrier in the presence of an ac field, obtained in a semiclassical approximation \cite{ivl86}.
We consider the limits of Eq.~\rqn{6.1} for low and high frequencies, which will be used below.

 With the help of Eqs.~\rqn{2.36}, \rqn{5.24}, \rqn{5.28}, and \rqn{5.29}, we obtain the following simplified expressions.
 For low frequencies
 \bes{6.2}
 \bea
&e^{-2C_0}\approx&\exp\left[-\frac{2}{3({\cal E}+F)}+
\frac{F\omega^2}{15({\cal E}+F)^4}\right],\nonumber\\
&&\omega\ll{\cal E}+F,
\ea{6.2a}
which agrees in the limit $\omega\rightarrow0$ with the exponential factor in the adiabatic expression \rqn{4.7} and provides the first $\omega$-dependent correction to the exponent.

For sufficiently high frequencies, expanding $C_0$ to the lowest orders in $\cal E$ and $F$ yields \cite{note3},
 \be
e^{-2C_0}\approx e^{-1/(2\omega)}\left(\frac{eF}{2\omega}
\right)^{2M},\ \ \omega\gg F,{\cal E}^{2/3}\ln\frac{\omega}{F};
 \e{6.2b}
 \ese
where
 \be
M=\frac{1}{2\omega}+\frac{F^2}{4\omega^3}-\frac{{\cal E}}{\omega^2}
\approx\frac{|E_0|+E_c}{\omega}.
 \e{6.3}
In the last equality, we took into account Eqs.~\rqn{5.79} and \rqn{6.14b} and the fact that $E_0=-1/2$ in the dimensionless units.
 Equation \rqn{6.2b} hints that for high frequencies the detachment is caused mainly by multiphoton absorption in the ac field \cite{note}. (See Sec.~\ref{VIB3} for more details.)

We turn now to the exact results.

\subsection{The limit of a weak static field}

For a sufficiently weak static field,
 \be
{\cal E}\ll F+\omega,
 \e{6.41}
the results obtained in Sec.~\ref{VC2} can be simplified with the help of the following formulas,
 \bea
&&b_0\approx2d_2\approx1/C_4\approx\sqrt{F^2+\omega^2},\
C_3\approx1,\ C_5\approx2C_0,\nonumber\\
&&\Delta_1^{-1}\approx(2/\omega)\,{\rm
arcsinh}(\omega/F)-2(F^2+\omega^2)^{-1/2},\nonumber\\
&&\Delta_2\approx2^{-1/3}{\cal E}^{2/3},\
z_0\approx\sqrt{F^2+\omega^2}/\omega^2.
 \ea{6.39}
Using Eqs.~\rqn{6.2} and \rqn{6.39}, we obtain that in the limit ${\cal E}\rightarrow 0$, expression \rqn{5.55} reduces to the well known result of detachment in an ac field
\cite{kel64,nik66,per66,per66a}, divided by two due to the present assumption (see Sec.~\ref{VA}) that we neglect the detachment occurring during the time intervals when the ac field is directed opposite to the static field.

\subsection{The total rate for sufficiently low frequencies}
 \label{VIB}

In this and following subsections, we provide simple analytical formulas for the total detachment rate in different regimes.

In the present subsection, we consider the case in which the spectrum \rqn{5.62} includes many peaks with comparable heights.
As shown below, this occurs for sufficiently low frequencies, namely, when either of the following conditions holds,
 \be
\omega\ll F\quad\text{or}\quad\omega\ll{\cal E}^{2/3}.
 \e{6.15}

 In this case, the partial probabilities $W_n$ slowly vary as a
function of $n$, and the sum \rqn{5.66} can be approximated by the
integral
 \be
W\approx\frac{1}{\omega}\int_{-\infty}^{\infty}dE\,W(E),
 \e{6.10}
where
 \be
W(E)=W_n|_{E_n\rightarrow E}.
 \e{6.8}
Using Eqs.~\rqn{6.8}, \rqn{5.62b}, and \rqn{5.64} in \rqn{6.10} and changing the order of integration, we integrate first over the energy with the help of the formula,
 \be
\int_{-\infty}^{\infty}d\xi\,{\rm Ai}^2(\xi)e^{a\xi}=\frac{e^{a^3/12}}
{2\sqrt{\pi a}}\quad(a>0),
 \e{6.9}
derived in Appendix \ref{B}.
 Performing the remaining integration over $u$ yields finally,
\be
 W=\frac{A^2\omega}{2\eta b_0}\sqrt{\frac{\pi}{c_0}}\,e^{-2C_0}.
 \e{5.68}
Equation \rqn{5.68} can be compared with Eq.~\rqn{6.1}.
In contrast to Eq.~\rqn{6.1} obtained with exponential accuracy, expression \rqn{5.68} provides the exact pre-exponential factor.

For low frequencies, Eq.~\rqn{5.68} can be simplified using Eqs.~\rqn{5.24}, \rqn{5.29a}, \rqn{5.31a}, and \rqn{6.2a},
 \bea
&W=&\frac{A^2}{2}\sqrt{\frac{3\pi}{F}}({\cal E}+F)^2
\exp\left[-\frac{2}{3({\cal E}+F)}\right.\nonumber\\
&&\left.+\frac{F\omega^2}{15({\cal E}+F)^4}\right],\quad
\omega\ll{\cal E}+F.
 \ea{6.11}
This expression coincides with the adiabatic formula \rqn{4.7}, if one disregards the second term in the exponent of Eq.~\rqn{6.11}.
Thus, this term provides an $\omega$-dependent correction to the adiabatic result \rqn{4.7}.
Note that the condition for this term to be significant coincides with the validity condition \rqn{5.38} of the present theory.
Although for
\be
 F\omega^2<({\cal E}+F)^4,
 \e{6.20}
strictly speaking, the present theory is not valid, we can conclude from Eq.~\rqn{6.11} by continuity considerations that in the region given in Eq.~\rqn{6.20} the total rate is described by the adiabatic result \rqn{4.7}, at least, when $\omega\ll{\cal E}+F$.

Let us obtain the validity conditions for the above results.
The spectrum contains many peaks with comparable heights if the characteristic width of the spectrum $\Delta_w$ as a function of $E_n'$ is much greater than the field frequency,
\be
\Delta_w\gg\omega.
 \e{6.34}
Analysis of Eq.~\rqn{5.62b} with the account of Eq.~\rqn{5.64} shows that, for sufficiently large and positive $E_n'$, the spectrum is cut off due to the factor exp$(-E_n'/\Delta_1)$, whereas for negative $E_n'$ the dependence of $W_n$ on $E_n'$ is given roughly by
\be
 W_n\propto\exp[|E_n'|/\Delta_1-(4/3)(|E_n'|/\Delta_2)^{3/2}],\quad
-E_n'\gg\Delta_2.
 \e{6.19}

As a result, we obtain that, for $\Delta_1\gg\Delta_2$, the distribution of $W_n$ has a maximum near $E_n'=0$ and practically vanishes for $E_n'\gg\Delta_1$ and also for $-E_n'\gg\Delta_2$, so that the effective width of the distribution of $W_n$ as a function of $E_n'$ is $\Delta_w=\Delta_1$.
Moreover, for $\Delta_1\ll\Delta_2$, Eq.~\rqn{6.19} implies that the distribution of $W_n$ as a function of $E_n'$ has a nearly Gaussian shape,
\be
 W_n\propto\exp[-(E_n'-E_{\rm max})^2/\Delta_G^2],\quad\Delta_1\ll\Delta_2.
 \e{6.21}
This distribution is centered at the negative energy,
$E_{\rm max}=-\Delta_2^3/(4\Delta_1^2)$, and has the effective width 
$\Delta_w=\Delta_G\equiv\sqrt{\Delta_2^3/\Delta_1}$.
The above results imply that for an arbitrary relation between $\Delta_1$ and $\Delta_2$ we have $\Delta_w=\max\{\Delta_1,\Delta_G\}$.

Consider two limits.
For low frequencies, $\omega\ll{\cal E}+F$, Eq.~\rqn{6.34} is fulfilled.
[Indeed, Eq.~\rqn{5.61} with the account of Eqs.~\rqn{5.31a} and \rqn{5.71} implies that $\Delta_1=3({\cal E}+F)^3/\omega^2$ and hence we obtain that $\omega\ll\Delta_1\le\Delta_w$.]
For high frequencies [$\omega\gg{\cal E}\ln(\omega/F),F$], Eq.~\rqn{5.61} with the account of Eqs.~\rqn{5.31b} and \rqn{5.71} implies that $\Delta_1=\omega/\ln[2\omega/(eF)],\ \Delta_2=({\cal E}^2/2)^{1/3}$, and hence $\Delta_G=\{{\cal E}^2\ln[2\omega/(eF)]/(2\omega)\}^{1/2}$.
Since now obviously $\omega>\Delta_1$, the condition \rqn{6.34} is equivalent to $\omega\ll\Delta_G$, which can be recast as $\omega\ll[{\cal E}^2\ln(\omega/F)]^{1/3}$.
Usually $[\ln(\omega/F)]^{1/3}\sim1$, and hence this factor can be omitted in the above inequality, yielding $\omega\ll{\cal E}^{2/3}$.
Thus, we obtain that Eq.~\rqn{6.34} holds under the condition that either $\omega\ll{\cal E}+F$ or $\omega\ll{\cal E}^{2/3}$; this condition is equivalent to Eq.~\rqn{6.15}.

\subsection{High frequencies}
 \label{VIB3}

Consider the region of sufficiently high frequencies,
\be
 \omega\gg F,{\cal E}^{2/3}\ln\frac{\omega}{F}.
 \e{6.22}
Now we can use Eq.~\rqn{6.2b}.
Moreover, in view of Eqs.~\rqn{5.61}, \rqn{5.70}, and \rqn{5.71}, now
\be
d_2\approx\omega/2,\quad
\Delta_2\approx({\cal E}^2/2)^{1/3}\equiv\omega_{\cal E},
 \e{6.18}
where $\omega_{\cal E}$ is the characteristic frequency for the Franz-Keldysh effect \cite{fra58,kel58,ans81,gar06}; in the usual units [cf.\ Eqs.~\rqn{4.3a} and \rqn{4.3c}]
\be
\omega_{\cal E}=\left(\frac{{\cal E}^2}{2\hbar m}\right)^{1/3}.
 \e{6.30}
In the present case, $d_2\gg\Delta_2$ [cf.\ Eqs.~\rqn{6.22} and \rqn{6.18}], which allows us to simplify Eq.~\rqn{5.64} in two overlapping regions.

 First, for $E_n'\gg\Delta_2$, the square of the Airy function in the integrand in Eq.~\rqn{5.64} can be approximated for negative values of the argument by its asymptotic expansion, the oscillations of which are smoothed out [cf.\ Eq.~\rqn{3.35b}], and for positive values of the argument by zero [cf.\ Eq.~\rqn{3.35a}].
 This yields
\be
 {\rm Ai}^2\left(\frac{d_2u-E_n'}{\Delta_2}\right)\approx \frac{{\cal
E}^{1/3}\theta(E_n'-d_2u)}{2^{7/6}\pi\sqrt{E_n'-d_2u}},
 \e{6.23}
where $\theta()$ is the step function.
 On inserting Eq.~\rqn{6.23} into \rqn{5.64} and performing the
integration, we obtain from Eqs.~\rqn{5.62b}, \rqn{6.2b}, \rqn{6.3}, \rqn{6.39}, and \rqn{6.18} that for sufficiently high-energy peaks, $E_n'\gg\omega_{\cal E}$, i.e., $n-M\gg\omega_{\cal E}/\omega$, the partial detachments rates are
 \be
W_n=2A^2\omega^{3/2}F_1\left(\sqrt{2(n-M)}\right)
e^{-1/(2\omega)}\left(\frac{eF}{2\omega}\right)^{2n},
 \e{6.24a}
where $F_1(\xi)$ is Dawson's integral \cite{abr64},
\be
 F_1(\xi)=e^{-\xi^2}\int_0^\xi e^{u^2}\,du\approx
 \left\{\begin{array}{ll}
\xi,&\ \xi\ll 1\\
1/(2\xi),&\ \xi\gg 1
\end{array}\right.
 \e{6.25}
The maximum of $F_1(\xi)$ is $F_1(0.92)=0.54$.

Second, for peaks with low and and moderately high energies, $E_n'\le0$ or $0<E_n'\ll d_2$, the exponential in the integrand in \rqn{5.64} can be neglected, whereas the approximate expressions for the parameters used in the derivation of Eq.~\rqn{6.24a} [i.e., Eqs.~\rqn{6.2b}, \rqn{6.39}, and \rqn{6.18}] are still applicable now.
As a result, Eq.~\rqn{5.62b} becomes, for $n-M\ll1$ or $n<M$,
 \be
W_n=2^{3/2}A^2\omega\sqrt{\omega_{\cal E}}\,
f\left(\frac{E_n'}{\omega_{\cal E}}\right)e^{-1/(2\omega)}
\left(\frac{eF}{2\omega}\right)^{2n},
 \e{6.24b}
where
\be
 f(\xi)=\pi\int_{-\xi}^\infty {\rm Ai}^2(\zeta)\,d\zeta
 =\pi\{[{\rm Ai}'(-\xi)]^2+\xi{\rm Ai}^2(-\xi)\}.
 \e{6.38}
Here ${\rm Ai}'(\xi)$ is the derivative of the Airy function
\cite{abr64}.
The derivation of the second equality in Eq.~\rqn{6.38} is given in Ref.~\cite{man00}.
 Using asymptotic expansions for the Airy function and its
derivative \cite{abr64} [cf.\ Eqs.~\rqn{3.35}], we obtain
\bes{6.37}
 \bea
 &&f(\xi)\approx\sqrt{\xi}+
 \frac{\cos(4\xi^{3/2}/3)}{4\xi},\ \ \xi\gg 1\label{6.37a}\\
 &&f(\xi)\approx\frac{\exp(-4|\xi|^{3/2}/3)}{16|\xi|^{3/2}},
 \ \ \ \ \ \ -\xi\gg 1.
 \ea{6.37b}
 \ese
Moreover, Eq.~\rqn{6.38} and the formula 10.4.5 in Ref.~\cite{abr64} imply that
\be
 f(0)=\frac{\pi}{3^{2/3}\Gamma^2(1/3)}\approx0.210,
 \e{6.50}
where $\Gamma(\xi)$ is the gamma function \cite{abr64}.
The first equality in Eq.~\rqn{6.38} implies that $f(\xi)$ is a positive, monotonically increasing function.

In the region \rqn{6.22}, detachment is a typical multiphoton process.
The behavior is similar to that in the absence of the static field \cite{kel64}, but there are also important features peculiar for the present case of a non-zero static field.
In particular, multiphoton detachment depends significantly on the threshold given by
 \be
E_{\rm th}=M\omega=|E_0|+E_c=|E_0|+U_p-{\cal E}/\omega.
 \e{6.31}
The threshold is increased by the value of the ponderomotive energy $U_p$, Eq.~\rqn{2.23}, which is a well known feature of above-threshold ionization \cite{kel64}.
However, in the present case the threshold is also decreased by a quantity proportional to the static field.
Moreover, in contrast to the case of a zero static field, now the threshold is not sharp, being smoothed by the quantity of the order of $\omega_{\cal E}$.

Equations \rqn{6.24a} and \rqn{6.24b} together with Eqs.~\rqn{6.25} and \rqn{6.37b} imply that $W_n$ depends very strongly on $n$.
The quantities $W_n$ decrease rapidly with $n$ increasing above the threshold, $n-M\ge1$, and practically vanish significantly below the threshold, $M-n\gg\omega_{\cal E}/\omega$.
However, $W_n$ does not vanish for $M-n\alt\omega_{\cal E}/\omega$, so that the effective threshold $E_{\rm th}'$ is actually lower than $E_{\rm th}$ by an amount of $\sim\omega_{\cal E}$.

In a typical case, $W_n$ is maximal for $n=[M]+1$, which is the minimal number of photons required to excite the atom above the threshold.
 (Here $[M]$ is the integer part of $M$.)
Moreover, since the higher peaks decrease with $n$ approximately as a geometric progression with the common ratio equal to $(eF)^2/(2\omega)^2\ll1$ [cf.\ Eq.~\rqn{6.24a}], $W_{[M]+1}$ is much larger than the sum of the other quantities $W_n$, and hence the total detachment probability
 \be
W\approx W_{[M]+1}.
 \e{6.32}

This relation can be incorrect only in relatively rare cases in which there is a peak sufficiently close to the threshold, so that the peak number $n_0$ satisfies $|n_0-M|\ll1$ (where $n_0$ can be less or greater than $M$).
Then the peaks $n_0$ and $n_0+1$ can be of the same order of magnitude, so that
 \be
W\approx W_{n_0}+W_{n_0+1}.
 \e{6.28}

\begin{figure}[htb]
\begin{center}
\includegraphics[width=8cm]{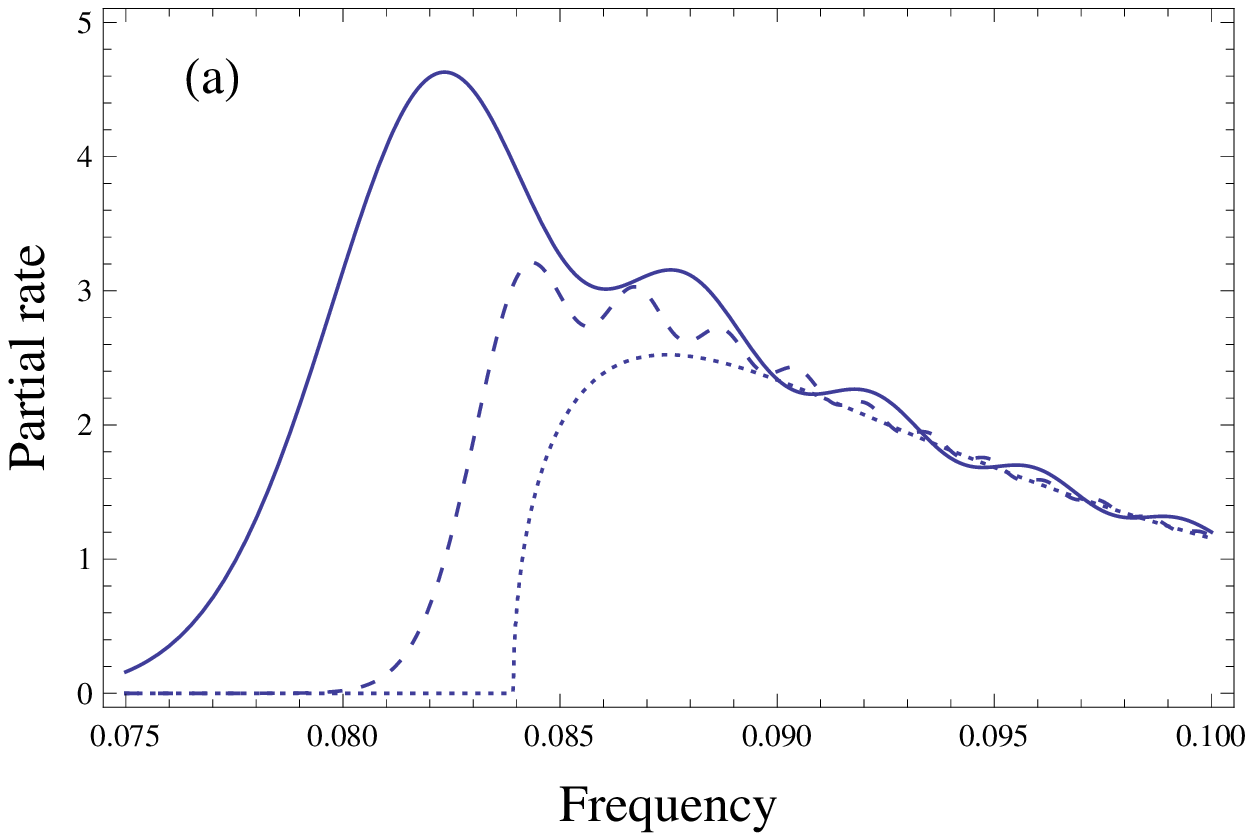}
\includegraphics[width=8cm]{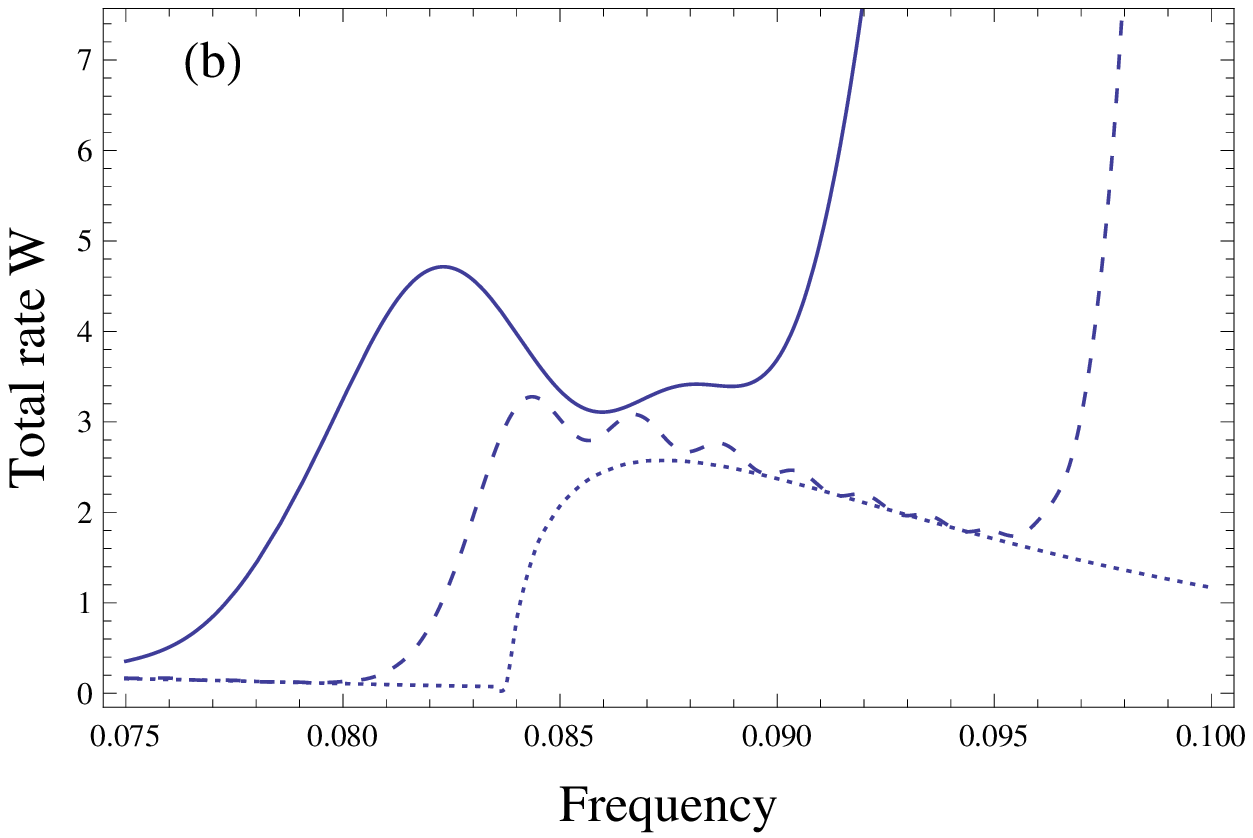}
\end{center}
\caption{The partial rate, $W_6$, (a) and the total rate $W$ (b) in units of $2\times10^{-15}|E_0|/\hbar$ versus the frequency $\omega$ for $F=0.01$. The solid lines: ${\cal E}=0.003$, the dashed lines: ${\cal E}=0.001$, the dotted lines: ${\cal E}=0$. The quantities $\omega,\ F$, and ${\cal E}$ are given in the dimensionless units.
}
 \label{f2}\end{figure}

Figure \ref{f2}(a) shows the partial rate $W_6$ as a function of the frequency $\omega$ for two nonzero values of $\cal E$ and in the absence of the static field.
These plots show that the partial rate peaks when $\omega$ is near the threshold, i.e., when $|n_0-M|\ll1$.
Therefore it is of interest to discuss the behavior near the threshold in more detail.
In this case, detachment is described by Eq.~\rqn{6.24b}, which implies that detachment depends significantly on the static field.
The static field results in several effects.
First, the threshold energy is decreased in the presence of the static field [see Eq.~\rqn{6.31}].
Second, the threshold is smoothed out.
As a result, detachment is possible below the threshold [see Eqs.~\rqn{6.24b} and \rqn{6.37b}] and is nonzero at the threshold [cf.\ Eq.~\rqn{6.50}].

Third, sufficiently above the threshold, the partial detachment probability, corresponding to the peak near the threshold, oscillates as a function of $E_{n_0}'$.
Indeed, Eqs.~\rqn{6.24b} and \rqn{6.37a} imply that
 \bea
&W_{n_0}&=2^{3/2}A^2\omega e^{-1/(2\omega)}\left(\frac{eF}{2\omega}\right)^{2n_0}\left\{\sqrt{E_{n_0}'}\right.\nonumber\\
&&\left.+\frac{\omega_{\cal E}}{4\sqrt{E_{n_0}'}}
\cos\left[\frac{4}{3}\left(\frac{E_{n_0}'}{\omega_{\cal E}}\right)^{3/2}\right]\right\},\nonumber\\
&&\quad\omega_{\cal E}\ll E_{n_0}'\ll\omega.
 \ea{6.51}
The second term in Eq.~\rqn{6.51} demonstrates oscillations of $W_{n_0}$ as a function $E_{n_0}'=\omega(n_0-M)$.
The quantity $E_{n_0}'$ can be changed, and hence the oscillations of $W_{n_0}$ can be observed on varying any of the quantities $\omega$, $F$, and $\cal E$.
The oscillations decrease with $E_{n_0}'$; the amplitude of the oscillations and the intervals between them increase with $\cal E$.
Note that $W_{n_0}$ decreases with $\omega$ sufficiently above the threshold [cf.\ Fig.~\rqn{f2}(a)]; this feature is determined by the factor before the braces in Eq.~\rqn{6.51}, while all other frequency-dependent factors in Eq.~\rqn{6.51} increase with $\omega$.

Since the partial rate $W_{n_0}$ can provide a significant or even dominant contribution to the total rate \rqn{6.28}, the above features of the near-threshold behavior generally apply not only to $W_{n_0}$ but also to $W$, see Fig.~\rqn{f2}(b).
The above near-threshold behavior is reminiscent of that obtained in the case of the Franz-Keldysh effect \cite{fra58,kel58,ans81,gar06} or one-photon detachment of ions in an electric field \cite{bry87,du88}.
There are, however, also features peculiar to the present case.
In particular, in contrast to the Franz-Keldysh effect, in the present case, oscillations of the detachment rate can occur, as mentioned above, not only as a function of the laser frequency, but also as a function of the laser-field intensity or the static-field strength.

\section{Application to electron emission}
\label{VII}

Consider electron emission from a metal or semiconductor surface or from photosynthetic bio-complexes due
to a static field perpendicular to the surface, under conditions
typical for scanning tunneling microscopy \cite{sca93,Lukins1,Lukins2}.
 Let $V_0$ be the tip (positive) potential with respect to the
surface and $d$ be the tip-surface separation.
 Then the static-field strength ${\cal E}_f=V_0/d$.
 In view of Eq.~\rqn{4.3c}, the dimensionless quantity ${\cal E}=|e_0|{\cal E}_f/F_a$ or, in view of Eq.~\rqn{4.3c},
\be
{\cal E}=\frac{0.0974}{d|E_0|^{3/2}}V_0,
 \e{7.2}
where $d$, $E_0$, and $V_0$ are measured in nm, eV and V,
respectively.
 For example, for $d=1$ nm, $E_0=-5$ eV, and $V_0=1$ V Eq.~\rqn{7.2}
yields ${\cal E}=0.0087$.

Assume that a laser (or ac) field perpendicular to the surface is superimposed on the static field.
As follows from the above theory, the laser field affects electron emission only if the field intensity is sufficiently high.
 Let us estimate the minimal strength of the laser field
required to affect significantly electron emission.
 The lower limit on the ac-field amplitude, at least, for low frequencies, $\omega\ll F+{\cal E}$, is implied by Eq.~\rqn{4.11}, $F\agt{\cal E}^2$, or
 \be
F_u\agt{\cal E}_a{\cal E}^2,
 \e{7.5}
where $F_u$ is the laser electric field amplitude in V/cm, the subscript $u$ indicates that the quantity is measured in usual units, and
 \be
{\cal E}_a=F_a/|e_0|=1.03\times 10^8|E_0|^{3/2}\text{ V/cm}.
 \e{7.6}
Here $E_0$ is in units of eV, and in the last equality we used Eq.~\rqn{4.3c}.
For example, for $E_0=-5$~eV, we obtain ${\cal E}_a=1.1\times10^9$~V/cm.
If also ${\cal E}=0.01$, Eq.~\rqn{7.5} yields $F_u\agt10^5$ V/cm, which corresponds to the laser intensity
$I_L=\epsilon_0cF_u^2/2\agt1.7\times10^7$~W/cm$^2$, where $\epsilon_0$ is the vacuum permittivity.
Note that the minimum laser intensity in the latter condition increases with $|E_0|$ as $|E_0|^3$ [cf.\ Eqs.~\rqn{7.5} and \rqn{7.6}].

Consider the field frequency.
The present theory implies that electron-emission rate increases with the laser-field frequency when the latter is sufficiently high.
Let us estimate the minimal frequency required to affect significantly electron emission.
 Equation \rqn{4.3a} implies that the field frequency measured in hertz, $\omega_u$, is related to the dimensionless frequency $\omega$ by
\be
\omega_u=4.84\times 10^{14}|E_0|\omega\text{ Hz}.
 \e{7.3}
As follows from Eq.~\rqn{6.11}, the field frequency significantly affects the emission rate for
 \be
\omega>\frac{({\cal E}+F)^2}{\sqrt{F}}
 \e{7.4}
[cf.\ also Eq.~\rqn{5.38}].
 If ${\cal E}=F=0.01$ and $E_0=-5$ eV, then Eqs.~\rqn{7.4} and \rqn{7.3} imply that the frequency affects the transition rate for
$\omega_u>10^{13}$ Hz.
 In view of the validity condition \rqn{5.25}, we obtain in this example that in the frame of the present theory, the frequency affects detachment in the interval
$10^{13}<\omega_u\alt 2\times 10^{14}$ Hz or $1.5\alt\lambda<30$ $\mu$m, where $\lambda$ is the wavelength of the laser field.

\section{Conclusion}
\label{VIII}

In this paper we have studied multiphoton electron detachment from atoms or  negative ions by a sum of a static and a laser (or, more generally, ac) fields.
Though, strictly speaking, the present theory holds for electron detachment from negative ions, the main qualitative features of detachment revealed here should be applicable also to ionization of neutral atoms.
We have developed an analytical theory of this phenomenon and derived simple formulas valid in a broad range of the parameters of the problem.
In particular, we have obtained that the energy spectrum of the electrons consists of narrow equidistant peaks separated by $\hbar\omega$.

We have identified, at least, two physically different regimes.
For low and moderately high frequencies, there are many spectral peaks of comparable size, and, as a result, the total rate smoothly depends on the frequency.
 For very high frequencies, the detachment is reminiscent of that
in the absence of the static field, including the appearance of the
energy threshold.
We have shown that the static field can significantly affect detachment for both low and high frequencies.
The effects of the static field include, in particular, oscillations of the energy spectrum of the products for low frequencies and the decrease and the smoothing of the threshold and oscillations of the total rate for high frequencies.
The results of the present work significantly clarify the physical picture of multiphoton electron detachment in the presence of a static electric field and provide a basis for further investigations in this field.

\acknowledgments
The work by G.P.B. was carried out under the auspices of the National Nuclear Security Administration of the U.S. Department of Energy at the Los Alamos National Laboratory under Contract No. DE-AC52-06NA25396.
A.G.K thanks the LANL for partial support at the initial stage of this work.

\appendix
\section{Average energy in different states of an electron in static and ac fields}
\label{A}

Here we calculate the quantum mechanical expectation values of the energy in different states describing a free electron simultaneously affected by static and ac fields [Eq.~\rqn{3.29'}].
More specifically, we shall consider the Volkov functions and the functions $|\psi_{E\vec{q}}(t)\rangle$.
We shall also obtain time averages of the expectation values of the energy over the electron oscillations.

The energy operator of an electron in a static field is given by [in the units \rqn{3.3}]
\be
H=\frac{\hat{\vec{P}}^2}{2}-\vec{\cal E}\cdot\vec{r}\equiv K+U,
 \e{A27}
where $K$ and $U$ are the kinetic and potential energies, respectively.
An ac field can change the electron energy with time.
It is of interest to obtain the instantaneous and time-averaged expectation values of the electron energy in the presence of the ac field, $F\sin\omega t$, for the basis states that are of relevance here.
The expectation value of the electron energy is given by the average of the operator $H$ [Eq.~\rqn{A27}] in the electron state.
It is convenient to work in the momentum representation, in which the kinetic and potential energies of an electron are, respectively,
 \be
K=\frac{P^2}{2},\quad
U=-i\vec{\cal E}\cdot\frac{\partial}{\partial\vec{P}}.
 \e{A2}

A difficulty in obtaining expectation values for basis states of the continuum is that such states are not normalizable to one [cf.\ Eqs.~\rqn{3.18} and \rqn{3.17}].
To overcome this difficulty, we consider expectation values for wave packets and obtain the result for a basis function in the limit when the width of the wave packet tends to zero.

\subsection{Volkov functions}
\label{A.1}

We begin with the Volkov functions.
The Volkov function in the momentum representation,
 \be
\psi_{\vec{p}_0}(\vec{P},t)=\int d\vec{r}\,
\psi_{\vec{p}_0}(\vec{r},t) e^{-i\vec{P}\cdot\vec{r}},
 \e{A3}
is given in the general case, in view of Eq.~\rqn{3.5}, by
\be
\psi_{\vec{p}_0}(\vec{P},t)=(2\pi)^3\delta(\vec{P}-\vec{P}_0(t)) e^{i\int_0^td\tau P_0^2(\tau)/2}.
 \e{A4}
Here
\be
 \vec{P}_i(t)=\vec{p}_i-\frac{e_0}{c}\vec{A}(t),
 \e{A5}
where $\vec{A}(t)$ is given in Eq.~\rqn{3.31}.
It is easy to see that the functions \rqn{A4} satisfy the orthonormality relation in the momentum representation [cf.\ Eq.~\rqn{3.18}],
 \be
\int\frac{d\vec{P}}{(2\pi)^3}\,\psi_{\vec{p}_0}^*(\vec{P},t) \psi_{\vec{p}_0'}(\vec{P},t)=(2\pi)^3\delta(\vec{p}_0-\vec{p}_0').
 \e{A7}

We consider a Gaussian wave packet in the momentum space,
 \be
\Psi_1(\vec{P},t)=\frac{(2\pi)^{-9/4}}{\Delta_p^{3/2}}\int d\vec{p}_0 \exp\left[-\frac{(\vec{p}_0-\vec{p}_1)^2}{4\Delta_p^2}\right] \psi_{\vec{p}_0}(\vec{P},t).
 \e{A6}
Here $\vec{p}_1$ is the average value of the vector $\vec{p}_0$, whereas $\Delta_p$ is the standard deviation of the components of $\vec{p}_0$.
In the limit $\Delta_p\rightarrow0$, the function in Eq.~\rqn{A6} becomes proportional to the Volkov function $\psi_{\vec{p}_1}(\vec{P},t)$.
It is easy to see that, in view of Eq.~\rqn{A7}, the state \rqn{A6} is normalized to one,
 \be
\int\frac{d\vec{P}}{(2\pi)^3}\,|\Psi_1(\vec{P},t)|^2=1.
 \e{A8}
On inserting Eq.~\rqn{A4} into Eq.~\rqn{A6}, we obtain $\Psi_1(\vec{P},t)$ in an explicit form,
 \bea
&\Psi_1(\vec{P},t)=&\frac{(2\pi)^{3/4}}{\Delta_p^{3/2}} \exp\left\{-\frac{[\vec{P}-\vec{P}_1(t)]^2}{4\Delta_p^2}\right.
\nonumber\\
&&\left.-\frac{i}{2}\int_0^td\tau\left[\vec{P}+\frac{e}{c}\vec{A}(t)
-\frac{e}{c}\vec{A}(\tau)\right]^2\right\},\quad
 \ea{A9}
where $\vec{P}_1(t)$ is given by Eq.~\rqn{A5}.

The expectation value of the energy in the state \rqn{A9} is
 \be
E_{\rm exp}=\langle H\rangle\equiv\int\frac{d\vec{P}}{(2\pi)^3}\,\Psi_1^*(\vec{P},t)H \Psi_1(\vec{P},t)=\langle K\rangle+\langle U\rangle,
 \e{A10}
where, in view of Eqs.~\rqn{A2} and \rqn{A9}, we obtain
\bes{A12'}
 \be
\langle K\rangle=\frac{\langle P^2\rangle}{2}= \frac{3\Delta_p^2}{2} +\frac{P_1^2(t)}{2},
 \e{A12}
 \bea
&\langle U\rangle&=-\vec{\cal E}\cdot\int_0^td\tau\left[\langle \vec{P}\rangle +\frac{e}{c}\vec{A}(t)-\frac{e}{c}\vec{A}(\tau)\right]\nonumber\\
&&=-\vec{\cal E}\cdot\int_0^td\tau\vec{P}_1(\tau).
 \ea{A13}
\ese
Here in the last equality we took into account the equality $\langle \vec{P}\rangle=P_1(t)$ and Eq.~\rqn{A5}.

The expectation value of the energy for the Volkov function is obtained in the limit $\Delta_p\rightarrow0$.
Then, on dropping the subscript 1 and taking into account Eqs.~\rqn{3.6} and \rqn{3.31}, we obtain from Eqs.~\rqn{A12'} that for the Volkov function $|\psi_{\vec{p}}(t)\rangle$ the mean kinetic and potential energies are
 \bea
&&\langle K\rangle=\frac{P^2(t)}{2} =\frac{1}{2}\left(\vec{p}+\vec{\cal E}t-\frac{\vec{F}}{\omega}\cos\omega t\right)^2,\nonumber\\
&&\langle U\rangle=-\vec{p}\cdot\vec{\cal E}t-\frac{{\cal E}^2t^2}{2} +\frac{\vec{\cal E}\cdot\vec{F}}{\omega^2}\sin\omega t.
 \ea{A14}
Equation \rqn{A14} shows that the magnitudes of the mean kinetic and potential energies increase as $t^2$.
However, in calculating the expectation value of the total energy $E_{\rm exp}=\langle K\rangle+\langle U\rangle$, the terms proportional to $t^2$ and $t$ cancel, and we obtain
 \bea
&E_{\rm exp}=&\frac{p^2}{2}-\frac{\vec{p}\cdot\vec{F}}{\omega}\cos\omega t+\frac{F^2}{2\omega^2}\cos^2\omega t\nonumber\\
&&+\vec{\cal E}\cdot\vec{F}\left(\frac{\sin\omega t}{\omega^2}-\frac{t\cos\omega t}{\omega}\right).
\ea{A15}
Finally, averaging $E_{\rm exp}$ over the electron oscillations, we obtain the simple result,
 \be
\bar{E}_{\rm exp}=\frac{p^2}{2}+U_p.
 \e{A16}

Consider two important special cases.
First, for the familiar Volkov functions describing an electron in a sinusoidal field, the above formulas hold with $\vec{\cal E}=0$ and $U=0$.
As a result, Eq.~\rqn{A14} implies that
 \be
E_{\rm exp}=\langle K\rangle =\frac{1}{2}\left(\vec{p}-\frac{\vec{F}}{\omega}\cos\omega t\right)^2.
 \e{A17}
The time-averaged expectation value of the energy is again given by Eq.~\rqn{A16}.

Second, for the Volkov functions describing an electron in a static uniform field, we have $\vec{F}=0$.
Then Eqs.~\rqn{A14} and \rqn{A15} yield
 \be
\langle K\rangle =\frac{(\vec{p}+\vec{\cal E}t)^2}{2},\quad
\langle U\rangle=-\vec{p}\cdot\vec{\cal E}t-\frac{{\cal E}^2t^2}{2},\quad
E_{\rm exp}=\frac{p^2}{2}.
 \e{A18}

\subsection{Functions $|\psi_{E\vec{q}}(t)\rangle$}
\label{A.2}

We shall consider the functions $|\psi_{E\vec{q}}(t)\rangle$ for the case of interest here---when the ac field is linearly polarized along the static field, i.e., when Eqs.~\rqn{3.1} and \rqn{3.2} hold.
The representation of these functions in momentum space is obtained as in Eq.~\rqn{A3}.
So, in view of Eq.~\rqn{3.19}, we obtain
 \be
 \psi_{E\vec{q}_0}(\vec{P},t)=\int_{-\infty}^\infty
\frac{dp_{0z}}{2\pi}h_{q_0}^*(p_{0z},E-U_p)\psi_{\vec{p}_0}(\vec{P},t),
 \e{A19}
where $\vec{p}_0=(p_{0x},p_{0y},p_{0z})$ and $\vec{q}_0=(p_{0x},p_{0y})$.
Inserting Eqs.~\rqn{3.16} and \rqn{A4} [where $\vec{P}_0(t)$ is defined in Eq.~\rqn{A5} with $\vec{A}(t)$ given in Eqs.~\rqn{3.33} and \rqn{3.32}] into Eq.~\rqn{A19} and performing the integration yields
 \bea
&\psi_{E\vec{q}_0}(\vec{P},t)&=\frac{(2\pi)^2}{{\cal E}^{1/2}} \exp\left\{\frac{i\tilde{P}_z(t)}{\cal E} \left[\tilde{E}-\frac{\tilde{P}_z^2(t)}{6}\right]\right.\nonumber\\
&&\left.-\frac{i}{2}\int_0^td\tau\tilde{P}_z^2(t,\tau)-\frac{iQ^2t}{2}\right\}\delta(\vec{Q}-\vec{q}_0),\quad\quad
 \ea{A20}
where
$\tilde{E}=E-U_p-Q^2/2,\ \tilde{P}_z(t)=P_z+(e_0/c)A(t)$, $\tilde{P}_z(t,\tau)=\tilde{P}_z(t)-(e_0/c)A(\tau)$, and $\vec{Q}=(P_x,P_y)$.
The function \rqn{A20} can be obtained also by a direct solution of the Schr\"{o}dinger equation \cite{gao90}.

Consider a wave packet in the momentum space with a Gaussian distribution of the quantities $E$ and $\vec{q}_0$,
 \bea
&\Psi_2(\vec{P},t)&=\frac{(2\pi)^{-7/4}}{\Delta_E^{1/2}\Delta_p}\int_{-\infty}^\infty dE\int d\vec{q}_0 \exp\left[-\frac{(E-E_1)^2}{4\Delta_E^2}\right.\nonumber\\ &&\left.-\frac{(\vec{q}_0-\vec{q}_1)^2}{4\Delta_p^2}\right] \psi_{E\vec{q}_0}(\vec{P},t).
 \ea{A21}
Here $E_1$ and $\vec{q}_1$ are the average values of $E$ and $\vec{q}_0$, respectively, and $\Delta_E$ ($\Delta_p$) is the standard deviation of $E$ (of each component of $\vec{q}_0$).
In the limit $\Delta_E,\Delta_p\rightarrow0$, the function in Eq.~\rqn{A21} becomes proportional to $\psi_{E_1\vec{q}_1}(\vec{P},t)$.
Using Eq.~\rqn{3.17}, it is easy to show that the wave function in Eq.~\rqn{A21} satisfies the normalization condition given by Eq.~\rqn{A8} where $\Psi_1(\vec{P},t)$ is replaced by $\Psi_2(\vec{P},t)$.
Inserting Eq.~\rqn{A20} into Eq.~\rqn{A21} and performing the integration yields
 \bea
&\Psi_2(\vec{P},t)&=\frac{2^{5/4}\pi^{3/4}\Delta_E^{1/2}}{\Delta_p{\cal E}^{1/2}}\exp\left\{-\frac{\Delta_E^2\tilde{P}_z^2(t)}{{\cal E}^2}\right.\nonumber\\
&&\left.-\frac{(\vec{Q}-\vec{q}_1)^2}{4\Delta_p^2}+\frac{i\tilde{P}_z(t)}{\cal E} \left[E_1-U_p-\frac{Q^2}{2}-\frac{\tilde{P}_z^2(t)}{6}\right]\right.\nonumber\\
&&\left.-\frac{i}{2}\int_0^td\tau\tilde{P}_z^2(t,\tau)-\frac{iQ^2t}{2}\right\}.
 \ea{A22}

Let us obtain the expectation values of the kinetic and potential energy in the state \rqn{A22}.
Taking into account the expressions for the kinetic and potential energy given in Eq.~\rqn{A2}, where now $U$ has the form $-i{\cal E}\frac{\partial}{\partial P_z}$, we obtain
 \bea
&&\langle K\rangle=\frac{{\cal E}^2}{8\Delta_E^2}+\frac{e_0^2A^2(t)}{2c^2} +\Delta_p^2+\frac{q_1^2}{2},\nonumber\\
&&\langle U\rangle=E_1-U_p-\Delta_p^2-\frac{q_1^2}{2}- \frac{{\cal E}^2}{8\Delta_E^2}\nonumber\\
&&\quad\quad+\frac{e_0{\cal E}}{c}\int_0^td\tau A(\tau).
 \ea{A23}
Both $\langle K\rangle$ and $\langle U\rangle$ diverge in the limit $\Delta_E\rightarrow0$, since they contain terms $\sim\Delta_E^{-2}$.

However, the expectation value of the total energy $E_{\rm exp}=\langle K\rangle+\langle U\rangle$ is independent of both $\Delta_E$ and $\Delta_p$.
Dropping the subscript 1, we obtain that the expectation value of the energy in the state $|\psi_{E\vec{q}}(t)\rangle$ is
 \be
E_{\rm exp}=E-U_p+\frac{e_0^2A^2(t)}{2c^2}+\frac{e_0{\cal E}}{c}\int_0^td\tau A(\tau)
 \e{A24}
or, taking into account Eq.~\rqn{3.32},
 \be
E_{\rm exp}=E-U_p+\frac{F^2}{2\omega^2}\cos^2\omega t +\vec{\cal E}\cdot\vec{F}\left(\frac{\sin\omega t}{\omega^2}-\frac{t\cos\omega t}{\omega}\right).
 \e{A25}
Finally, the average of this quantity over the electron oscillations yields
 \be
\bar{E}_{\rm exp}=E.
 \e{A26}

\section{Calculation of the integral in Eq.~\rqn{6.9}}
\label{B}

To calculate the integral in Eq.~\rqn{6.9}, we use the differential equation for $w(\xi)={\rm Ai}^2(\xi)$ \cite{abr64},
 \be
w'''-4\xi w-2w=0.
 \e{B1}
Let us denote
 \be
J=\int_{-\infty}^{\infty}d\xi\,{\rm Ai}^2(\xi)e^{a\xi}\quad(a>0).
 \e{B2}
Multiplying the both sides of Eq.~\rqn{B1} by $e^{a\xi}$ and integrating each term over $\xi$, using the rule of integration by parts for the first two terms, we obtain the differential equation for $J$,
 \be
\frac{dJ}{da}=\left(\frac{a^2}{4}-\frac{1}{2a}\right)J.
 \e{B3}
Equation \rqn{B3} can be easily solved, yielding
 \be
J=\frac{Ce^{a^3/12}}{\sqrt{a}},
 \e{B4}
where $C$ is an unknown constant.

To obtain $C$, we note that Eq.~\rqn{B4} implies that $J$ diverges in the limit $a\rightarrow0$, so that
 \be
a^{1/2}J\rightarrow C,\quad a\rightarrow0.
 \e{B5}
The asymptotic equations \rqn{3.35} imply that the divergent term in Eq.~\rqn{B2} arises due to the integration over large negative values of $\xi$, i.e., for $a\rightarrow0$ [cf.\ Eq.~\rqn{3.35b}]
\be
J\approx\int_{-\infty}^{-\xi_0}d\xi\,\frac{e^{a\xi}}{2\pi\sqrt{|\xi|}}\left[1-\sin\left(\frac{4}{3}|\xi|^{3/2}\right)\right],
\e{B6}
where $\xi_0$ is any number such that $\xi_0\gg1$.
The replacement of $\xi_0$ by 0 in Eq.~\rqn{B6} changes $J$ by a term which is finite for $a\rightarrow0$.
Moreover, the integration of the oscillating term in Eq.~\rqn{B6} can be shown to yield the result which is finite for $a\rightarrow0$.
Terms, which are finite for $a\rightarrow0$, are of a higher order than the terms $\sim a^{-1/2}$ and hence can be neglected to a first approximation.
Therefore, in the limit $a\rightarrow0$, Eq.~\rqn{B6} implies that
\bea
&a^{1/2}J\rightarrow &a^{1/2}\int_{-\infty}^{0}d\xi\,\frac{e^{a\xi}}{2\pi\sqrt{|\xi|}}
=\frac{a^{1/2}}{\pi}\int_0^\infty d\zeta\,e^{-a\zeta^2}\nonumber\\
&&=\frac{1}{2\sqrt{\pi}},
\ea{B7}
where the first integral was transformed by the change of variable $\xi=-\zeta^2$.
A comparison of Eqs.~\rqn{B5} and \rqn{B7} yields the equality $C=(2\sqrt{\pi})^{-1}$, inserting of which into Eq.~\rqn{B4} results in the formula \rqn{6.9}.


\begin{thebibliography}{99}
\bibitem{note1}
The Volkov functions for an electron in a static and an ac fields were obtained for the case of the velocity gauge in Ref. \protect\cite{nik72}.
\bibitem{note4}
Note that the definition \rqn{3.14} differs by the factor $\pi^{-1/2}$ from the definition of the Airy function in Ref.~\cite{lan77}.
\bibitem{note5}
This conclusion, in principle, may be not true for very high $\omega$.
A study of this possibility is out of the scope of the present paper.
Note, however, that taking the saddle points with odd $k$ into account can increase the detachment rate at most by a factor of two.
\bibitem{note3}
The validity condition of Eq.~\rqn{6.2b} is somewhat stricter than
that shown in Eq.~\rqn{6.2b}, since terms $\sim F^4/\omega^5$ have
been neglected in the exponent of Eq.~\rqn{6.2b}.
\bibitem{note}
The early work \protect\cite{aru70} provides formulas similar to our Eqs.~\rqn{6.2}.
However, the formulas in Eq.~\cite{aru70} miss some important terms, such as the $\omega$-dependent term in Eq.~\rqn{6.2a} and the contribution due to the ponderomotive energy [the first term in the first equality in Eq.~\rqn{6.3}], and contain some incorrect coefficients.
\end{thebibliography}
\end{document}